\documentclass[a4paper,fleqn,usenatbib]{mnras}
\usepackage{color}
\usepackage{ulem}
\usepackage{graphicx,url,amssymb,amsmath,longtable,rotating,units,wasysym,listings,bm}
\usepackage{natbib,verbatim,enumerate}

\newcommand\simlt{\lower.5ex\hbox{$\; \buildrel < \over \sim \;$}}
\newcommand\simgt{\lower.5ex\hbox{$\; \buildrel > \over \sim \;$}}

\title[Numerical simulations of AGN wind feedback]{Numerical simulations of AGN wind feedback on black hole accretion: probing down to scales within the sphere of influence} 
\author[M. Zeilig-Hess et al.]{Meir Zeilig-Hess$^1$\thanks{Email: meirzh10@gmail.com}, Amir Levinson$^{1,2}$ \& Ehud Nakar$^1$
\\
$1$ The Raymond and Beverly Sackler School of Physics and Astronomy, Tel Aviv University, Tel Aviv 69978, Israel\\
$2$Yukawa Institute for Theoretical Physics, Kyoto University, Oiwake-cho, Kitashirakawa, Sakyo-ku, Kyoto 606-8502, Japan\\
}

\begin{document}
\maketitle
	
\begin{abstract}
Several processes may limit the accretion rate onto a super-massive black hole (SMBH). Two processes that are commonly considered (e.g., for sub-grid prescriptions) are Bondi-Hoyle-Lyttleton accretion and the Eddington limit. A third one is AGN wind feedback.  It has been long suggested that such a wind feedback regulates the final SMBH mass, however, it has been shown recently that AGN winds can also regulate the average accretion rate at a level consistent with observations of high redshift AGNs.  In this paper we study the effect of wind feedback on the accretion rate using 2D, high resolution hydrodynamic simulations, that incorporate a self-consistent wind injection scheme and resolves the SMBH sphere of influence.  Two different cases are explored and compared: one in which the initial gas density is uniform, and one in which it has an isothermal sphere profile.  We also compare simulations with and without cooling. Our main finding is that for reasonable parameters, AGN feedback always limits the accretion rate to be far below the Bondi-Hoyle-Lyttleton limit. For typical wind parameters and a uniform ISM densities of $n\sim 1$ cm$^{-3}$, the accretion rate is found to be several orders of magnitudes smaller than that inferred in large samples of high redshift AGNs.  On the other hand, the accretion rate 
obtained for initially isothermal density profile is found to be consistent with the observations, particularly when cooling is included.  Furthermore, it roughly scales as $\sigma^5$ with the velocity dispersion of the bulge, in accord with the $M-\sigma$ relation
\end{abstract} 


\section{Introduction}

AGN winds have long been thought to constitute an important feedback mechanism that regulates the growth of
supermassive black holes (SMBHs) in the early universe, and affects the evolution of their host galaxies. 
 Hydrodynamical cosmological simulations that include AGN feedback \cite[e.g.,][]{diMatteo2005,robertson2006,sijacki2007,debuhr2011,vogelsberger2014,schaye2015,sijacki2015,dubois2016,weinberg2018}  cannot resolve the detailed physics of accretion onto the SMBH and must resort to sub-grid prescriptions, commonly based on Bondi-Hoyle-Lyttleton accretion models.
Recently, \cite{negri2017} made a detailed comparison study of various methods developed in the past two decades, in 
an attempt to elucidate how different assumptions affects the  resultant black hole accretion rate. 
Their analysis indicates a large variation 
in the  accretion rate (and other properties) between the different feedback models reported in the literature.  
In particular,  simulations that invoke  more realistic schemes of wind injection \citep[e.g.,][]{ostriker2010,choi2012,choi2014,ciotti2017,negri2017}
find substantially lower accretion rates.  However, those latter studies, while incorporating important processes such as cooling, star formation 
and supernovae feedback into the analysis,  do not elucidate the details of 
the interaction of the AGN wind with the ambient gas, as well as its dependence
on initial and boundary conditions and on grid resolution.   Other simulations \citep[e.g.,][]{nayakshin2012,wagner2013,bourne2015,zubovas2016}
while studying various aspects of AGN feedback on different scales (e.g., triggering star formation, ablating clouds in a two phase media) invoke a constant wind power 
and, therefore, are unable to directly model the feedback mechanism on the AGN wind, which is the main focus of this work.

It has been argued recently \cite[][hereafter LN18]{levinson2018} that various measurements of BH mass, accretion rate and
Eddington ratio in large samples of AGNs in the  redshift interval $0 \le z \le 7$, indicate a roughly constant accretion rate 
at redshifts $z > 2$, with a mean value of a few tens $M_\odot / yr$ \citep{kurk2007,willott2010,trakhtenbrot2017}, and a sharp 
decline with cosmic time below $z\simeq 2$ \citep{trakhtenbrot2011,trakhtenbrot2012}.  The inferred Eddington ratios of 
sources in the accretion plateau ($z \ge 2$)  are scattered between $0.1$ and $1$, with a mean at 0.3 roughly, indicating mildly 
sub-Eddington accretion by the SMBHs in this sample.  Based on these data LN18 argued that the accretion trend 
exhibited by the high redshift AGNs ($z>2$)
is consistent neither  with the infall rate of the gas in the hallo nor with the Eddington limit.    Furthermore, the inferred mass accretion rate 
seems to be considerably higher than that found in recent simulations that treat wind injection in a self-consistent manner. 
\citep[e.g.,][and references therein]{ciotti2017,negri2017}.

Motivated by these considerations, LN18 constructed a simple
analytic model for the interaction of an AGN wind with the galactic medium, in which the accretion rate is limited by  momentum balance between 
the wind and the infalling matter.   The tacit assumption underlying this model is that once the accretion rate exceeds this critical value, the shocked bubble created by the expanding wind (henceforth termed cocoon\footnote{We adopt the definition commonly used in the GRB literature.   Originally,  the term cocoon was coined to indicate the shocked wind (or jet) bubble alone.})
will push all the matter surrounding it, thereby completely halt accretion, chocking the wind.  As the wind weakens
accretion is resumed.   LN18 have shown that this intermittent wind injection process keeps the mean accretion rate roughly constant, at a level
consistent with the observations described above.     Once the expanding cocoon expels the 
entire gas in the bulge, black hole growth ceases.  This gives rise to an $M-\sigma$ relation \cite[][ and references therein]{kormendy2013}, in a manner similar to that proposed originally by \cite{silk1998} and later by \cite{king2003}, but with quantitative differences.  Previous analytic work \citep{king2003,king2010,zubovas2012,faucher2012,costa2014,king2015}, while studying various aspects of 
wind propagation and its interaction with the ambient medium, did not address the feedback on the wind injection. 

As mentioned in LN18, a caveat concerning their feedback scenario is the implicit assumption that accretion of shocked 
material is negligible.  It could well be that some filaments of shocked ambient matter produced by, e.g., Kelvin-Helmholtz and 
Rayleigh-Taylor instabilities \citep[e.g.,][]{nayakshin2012}, and/or dense matter accumulated around the equatorial plane, 
are being pushed in by the gravitational force and ultimately swallowed by the black hole.  This might alter the estimate 
of the regulated accretion limit derived in LN18.  Clumpy medium may also affect the feedback process \citep[e.g.,][]{nayakshin2012,wagner2013,bourne2014,costa2014}.
Additional assumption made in LN18 is that the wind is not highly collimated and that the ambient density is roughly spherical, as expected in high redshift bulges. If one of these assumptions is not satisfied, then the wind may escape the galaxy without depositing 
its entire energy in the bulge.   Wind collimation may also alter the shape and dynamics of the cocoon, and in particular the time it takes 
the shock to cross the bulge.

\begin{figure}
\includegraphics[width=1\columnwidth]{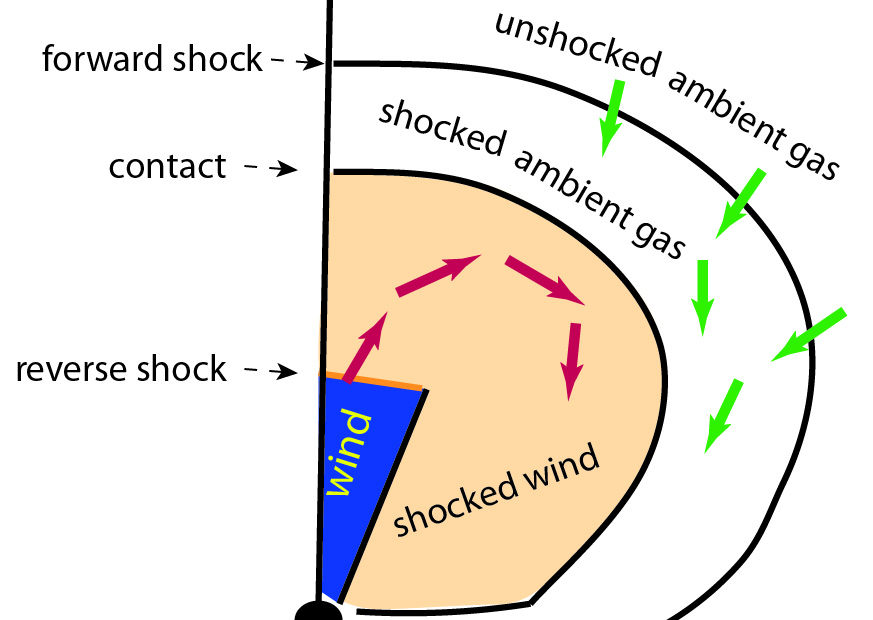}
\caption{Schematic illustration of the different flow components.  
The red and green arrows indicate the streamlines of shocked wind
material that crossed the reverse shock and shocked ambient  gas that crossed the forward shock, respectively.  
The shocked wind and ambient matter are separated by a contact surface.}
\label{app:fig:scheme}
\end{figure}

The main goal of this work is to study the hydrodynamics of wind feedback on the accretion process down to scales smaller than the radius of the SMBH sphere of influence. To do that, we perform high resolution 2D hydrodynamical simulations that resolve such scales and capture the essence of the interplay between the wind and the infalling galactic matter.
The injection of the wind is treated in our numerical model in a self-consistent manner, 
similar to the method employed by  \cite{ciotti2017} and \cite{negri2017}, as explained in detail below.
We also compare runs with vastly different density distributions, and show that it can greatly affect the accretion rate and
the feedback physics.  In particular, the density distribution adopted in the works hitherto cited cannot account for the high accretion 
rates measured at in samples of high redshift AGNs.  
A diagram showing the structure of the cocoon and the different
flow components is given in Fig. \ref{app:fig:scheme}. 

\section{Numerical scheme}
\label{sec:mdot}

The numerical model computes  the interaction of a wind ejected from the inner boundary of the simulation domain with 
infalling matter in a spheroidal galaxy.     
The protogalaxy is modelled as an isothermal sphere of dark matter, having a 
radius $R_b$ and a constant velocity dispersion $\sigma = 300\, \sigma_{300}$ km s$^{-1}$, that contains
gas of density $\rho_g$. 
The total mass of the dark matter halo
is related to its radius and velocity dispersion through $M_b = 2\sigma^2 R_b/G$.   The gravitational potential of the protogalaxy is taken to be
\begin{equation}
\Phi_b = 2\sigma^2 \ln(r/r_a),
\end{equation}
where 
\begin{equation}
r_a = \frac{G M_{BH}}{\sigma^2}\simeq 5 M_8 \sigma_{300}^{-2}\quad {\rm pc}
\label{eq:ra}
\end{equation}
is the sphere of influence of the SMBH and $M_{BH}=10^8 M_8 M_\odot$  its mass.  The gravitational potential contributed by the 
black hole can be expressed as
\begin{equation}
\Phi_{BH} = -\sigma^2 \frac{r_a}{r}.
\end{equation}
The net gravitational potential included in our simulations is the sum: $\Phi = \Phi_b + \Phi_{BH}$.   The characteristic 
free-fall time within $r_a$,
\begin{equation}
t_a=\frac{r_a}{\sigma} \simeq 2\times10^4 M_8 \sigma_{300}^{-3}\quad {\rm yr},
\end{equation}
is henceforth used as our reference time.
Since the primary goal of this paper is to study the interplay between the accreted  gas and the wind, 
treating feedback in a self-consistent manner, we set the rotational velocity of the gas in the galaxy to zero.
This tacitly assumes that angular momentum is unimportant on scales resolved by the simulation.   Well within the 
sphere of influence the centrifugal barrier will ultimately lead to formation of a disk around the SMBH, from which 
the putative wind in expelled.   

The simulation domain extends from some inner boundary, taken to lie within the sphere of influence, $r_{in} \ll r_a$, to the outer 
edge of the protogalaxy, $r_{out}=R_b$.  In the results presented below the inner boundary is at $r_{in}=0.1r_a$.  We  have also run cases 
with other values of $r_{in}$ and verified that 
the results are not significantly affected by the choice of $r_{in}$ provided it is much smaller than $r_a$.
The wind is injected from the inner boundary within two symmetric cones of opening angle $\theta_w$
above and below the equatorial plane.   The wind power $L_w$ is parametrized in terms of the efficiency  $\epsilon$ according 
to:  $L_w  = \epsilon \dot{M}_{BH} c^2$, where $\dot{M}_{BH} = \dot{M}_{in}-\dot{M}_{w}$, 
\begin{equation}
\dot{M}_{in}(t)=2\pi r_{in}^2 \int_{\theta_w}^{\pi-\theta_w}  \rho_g(t,r_{in},\theta) v_r(t,r_{in},\theta) \sin\theta d\theta
\label{eq:dotM_in}
\end{equation}
is the mass accretion rate at the inner boundary of the simulation domain, $v_r(t,r,\theta)$ is the local radial velocity of the 
accreted gas at time $t$, and 
\begin{equation}
\dot{M}_{w}(t)=4\pi r_{in}^2 \int_0^{\theta_w}  \rho_w(t,r_{in},\theta) v_w(t,r_{in},\theta) \sin\theta d\theta
\label{eq:dotM_w}
\end{equation}
is the wind's mass flux.   In the examples presented below the wind 
is injected uniformly (both $v_w$ and $\rho_w$) along the inner boundary, with  a constant (time independent) velocity $v_w$  and opening angle $\theta_w=45^\circ$ (as well as $\theta_w=30^\circ$ and $60^\circ$ in some runs).  The wind density is determined, at every time step, from the relation $\dot{M}_wv_w^2/2 =\epsilon\dot{M}_{BH}c^2=\epsilon(\dot{M}_{in}-M_w)c^2 $ and is time dependent. 
The wind's Mach number, ${\cal M}=v_w/c_s$, here $c_s=(\gamma p_w/\rho_w)^{1/2}$ and $\gamma=5/3$ is the adiabatic index, is taken 
to be large ( ${\cal M} =10^2$ in most examples).  We find that the results are practically independent of the 
choice of ${\cal M}$ as long as the wind is highly supersonic (${\cal M}>>1$).   
The rate at which the SMBH accretes mass can be expressed in terms of the mass inflow rate through the inner boundary, $\dot{M}_{in}$,
and the wind parameters $\epsilon$ and $v_w$, as:  $\dot{M}_{BH}=\dot{M}_{in}/(1+2\epsilon c^2/v_w^2)$.   It is seen that 
$(v_w/c)^2 <<\epsilon$ implies $\dot{M}_{in} \gg \dot{M}_{BH}$ which corresponds to wind ejection from large disk radii. 

The simulations were performed using version 4.0 of the PLUTO code \citep{mignone2007}.  
A 2D axisymmetric grid in spherical coordinates ($r,\theta$) is employed,
with a regular spacing of the $\theta$ grid and non-uniform spacing of the radial grid, that allows higher concentration 
of grid points in the inner region.   The radial grid is divided into two patches,  with a uniform spacing in the region $r_{in} < r <10 r_a$ 
and logarithmic spacing beyond $10 r_a$.   The uniform patch contains 1000 gridpoints (or a resolution of  $10^{-2} r_a$) 
and the logarithmic patch 600 gridpoints.  The $\theta$ grid consists of 200 gridpoins.
We use  axisymmetric boundary conditions on the $\theta$ boundary and open boundary conditions at  $r_{out}$, and 
at  $r_{in}$  outside the wind injection zone (i.e., at $\theta_w \le \theta \le \pi-\theta_w$)\footnote{In runs where wind injection 
is switched off we use the open boundary condition on the entire $r_{in}$ boundary.}.

 In our simulations we use a fiducial SMBH mass of $M_{BH}=10^8 M_\odot$.
 As will be shown below,  in the case of isothermal density profile the results are independent of the SMBH mass, and
 the Eddington ratio can be readily scaled.  
For this choice of density profile we find that in most cases the accretion into the SMBH is supercritical if $M_{BH}<10^8~M_\odot$,
and in some cases it is supercritical even at $M_{BH}\simlt 10^9~M_\odot$.
One might then naively expect that in reality the majority of the mass inflowing from the sphere of influence will be 
expelled from large disk radii, before reaching the SMBH, as models of radiatively inefficient accretion flows (RIAF) 
predict \citep[e.g.,][]{begelman2012}.   However, LN18   argued that the interaction of outflows expelled from large
disk radii during the supercritical accretion phase with the surrounding matter is likely to  lead to
accumulation of the unbound gas above the disk, that in turn
exerts pressure on the disk and forces the infowing matter to ultimately reach 
the inner disk regions, wherefrom the  fast winds responsible for the feedback are expelled. 
What is the actual outcome of supercritical accretion under such conditions is unclear at present.  
One can partially address this issue by choosing appropriate wind parameterization.  
The one employed above allows us to consider both, fast winds from the innermost disk radii during 
supercritical accretion, and slower wind from larger radii of the RIAF.   More precisely, since $\dot{M}_{BH}/\dot{M}_w=v_w^2/2\epsilon c^2$,
the fraction of $\dot{M}_{in}$ that is absorbed by the SMBH is controlled by this choice; for a given extraction efficiency $\epsilon$,
smaller $v_w$ implies smaller $\dot{M}_{BH}$.  This represents winds that emanate from larger disk radii with a smaller kinetic energy.

In the case of a uniform density medium (case B below) the accretion rate is always highly subcritical.

\begin{figure*}
\includegraphics[width=8cm]{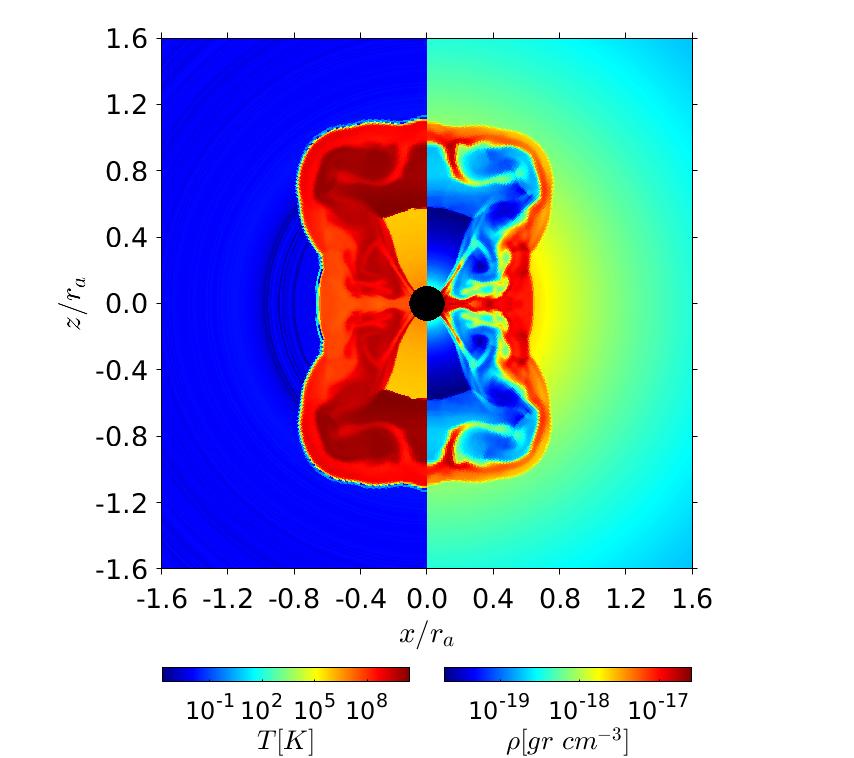} 
\includegraphics[width=8cm]{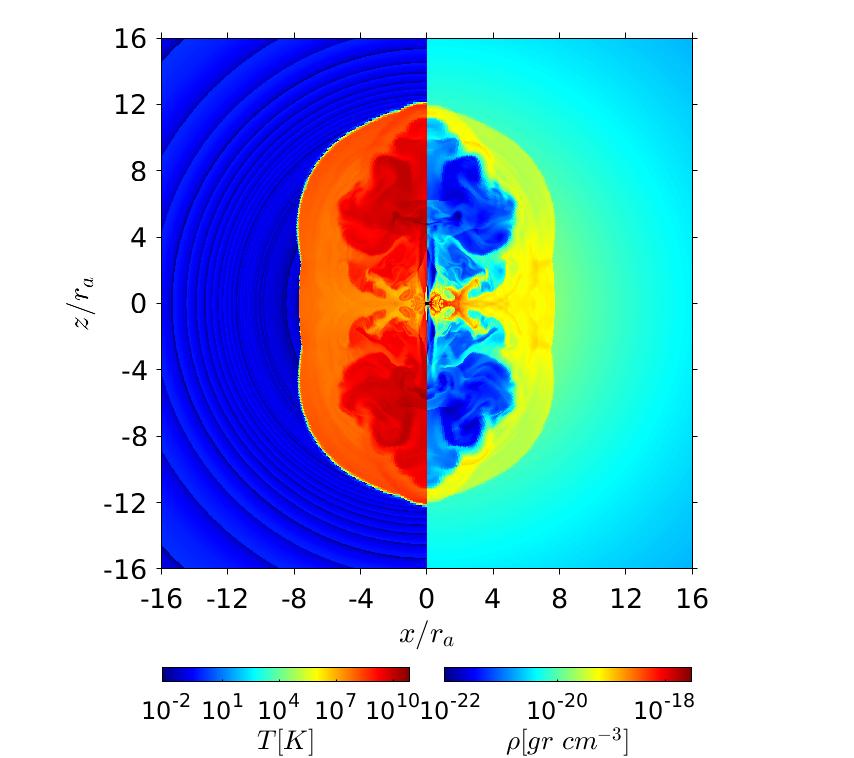} \includegraphics[width=8cm]{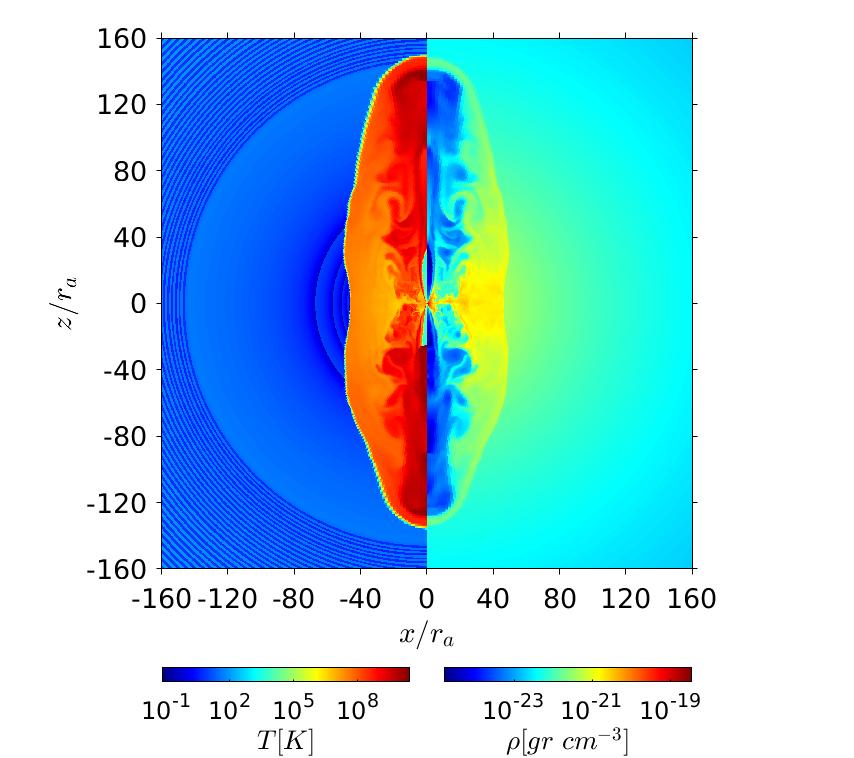}
\caption{Snapshots from the fiducial simulation in case A at time $t=0.1t_a$ (upper left panel), $1 t_a$ (upper right panel) and $t=10t_a$ (bottom panel),
showing density (right half) and temperature (left half) maps. 
Note the change in scales between the different images.}
\label{fig:density_iso}
\end{figure*}	

\section{Results}
\label{sec:results}
We performed two sets of numerical experiments.  In the first set (case A) the gas was taken to be initially at rest, with 
a density profile of an isothermal sphere, viz., $\rho_g(t=0,r)=f_g\sigma^2/2\pi G r^2 $, where $f_g=0.1$ is
the gas fraction in the protogalaxy. 
In the absence of an AGN wind the mass accretion rate is expected to quickly reach the dynamical limit
\begin{equation}
\dot{M}_{max}=4\pi\rho_gr^2\sigma\simeq 1.2\times10^4 f_g \sigma_{300}^3 \quad M_\odot yr^{-1},
\label{eq:dotM_iso}
\end{equation}
and remain constant afterwards \citep[e.g.,][]{king2010}.  
In general, accretion will commence once the gas cools sufficiently. 
If initially the gas is maintained at hydrostatic equilibrium, then its temperature is about $T\simeq m_p\sigma^2/k \simeq10^7 \sigma_{300}^2$ K.
Under these conditions the primary cooling mechanisms is free-free emission.   The free-free cooling time is estimated from Eq. (\ref{app:eq:ff_cooling})  to be
\begin{equation}
t_{ff}\simeq 10 \sigma_{300}^{-2} (f_g/0.1)^{-1}(r/r_a)^2(T/10^7\, K)^{1/2}\quad yr,
\label{eq:t_ff_iso}
\end{equation}
short compared with $t_a$ at  radii $r \simlt 30 r_a$\footnote{Note that with our normalization the Thomson depth 
is $\tau\simeq 5 (r/r_a)^{-1}$. Hence, optically thin cooling applies only at $r\simgt 5r_a$.}.  This  means that in practice, when accretion sets in the gas
is likely to be already cold.   
Hence, for practical purposes  the initial gas temperature can be 
taken to be small, $T_0<<m_p\sigma^2/k$.  We adopt this approach in the simulations with no cooling.
This, however, ignores the potential effect of cooling
on the shocked matter (as well as on the ISM), that might alter the evolution of the cocoon.   
Moreover, the shocked wind material 
may cool via inverse Compton scattering of the quasar radiation. 
Equation (\ref{app:eq:comp_cooling}) implies  
rapid cooling in the vicinity of $r_a$ for a luminosity near the Eddington limit, 
particularly in fast winds with $v_w/c > \sqrt{m_e/m_p}$, for  which the electrons in the 
shocked wind plasma are relativistic.   We shall come back to these  points in Sec. \ref{sec:cooling} below,
where the results for a run with strong cooling is discussed. 

As a test case, we performed a simulation with the wind injection switched off and compared the result to the 
analytic formula, Eq. (\ref{eq:dotM_iso}).   We find that after a short transient phase of about $2t_a$, the accretion rate 
saturates at a value which is larger by about 10\% than the analytic value (Fig \ref{fig:dotM_iso}).   This discrepancy is due to
our choice of the inner boundary.   Fixing the inner radius at $r_{in}= 0.01 r_a$ brings the numerical result to 
within 3\% of the analytic result.  However, we find that reducing $r_{in}$ requires 
higher resolution of the radial grid in order to avoid a numerical instability.    After experimenting with the location 
of the inner boundary  we concluded that $r_{in}=0.1 r_a$ is the optimal choice for our purposes. 

In the second set of experiments (case B) the initial gas density was taken to be uniform, $\rho_g(t=0,r)= m_p n_0$.  
Since the free fall velocity outside the sphere of influence is approximately constant, 
the density at a given radius $r$ at time $t > r/\sigma$ is expected to change according to $\rho(t,r)\simeq \rho_0(\sigma t/r)^2$ when
there is no feedback (i.e., when wind injection is switched off).
In particular, at the inner boundary $\rho(t)\simeq \rho_0(t/t_a)^2$ and $\dot{M}\simeq 4\pi \rho_0\sigma r_a^2 (t/t_a)^2$ 
(see appendix \ref{app:sec:accretion} for details).  Our test simulations with no wind injection reproduce this temporal 
accretion profile to a good accuracy (see Fig \ref{app:fig:Mdot}).  In this example we assumed that the gas is initially cold, 
that is, $T\ll m_p \sigma^2/k$.
If the gas is initially held at hydrostatic equilibrium, then accretion commences only after the gas sufficiently cools. 
The free-free cooling time is roughly  $t_{ff}\sim 10 n_0^{-1}$ Myr, with $n_0$ measured in c.g.s units (Eq. (\ref{app:eq:ff_cooling})). 
The Compton cooling time $t_c$ depends on the accretion rate.  For our fiducial simulation we find highly sub-Eddington accretion (see Sec. \ref{sec:caseB}),
hence we anticipate $t_c > t_{ff}$.

\begin{figure}
\includegraphics[width=8cm, height = 10 cm]{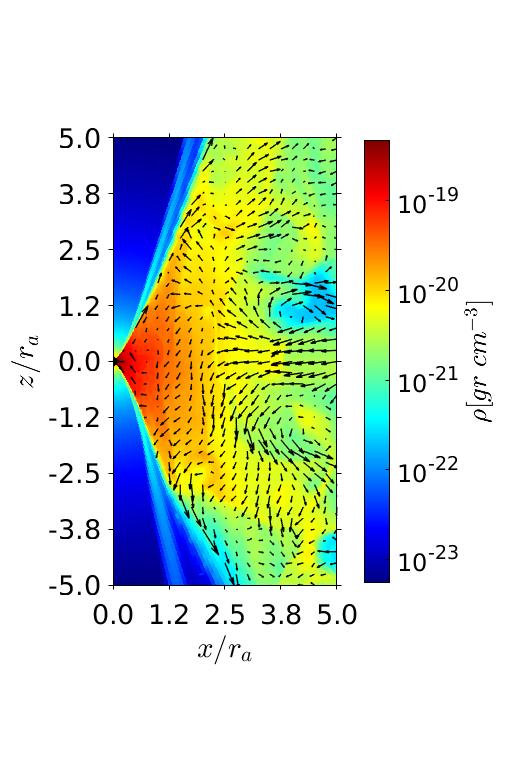}
\caption{Enlarged view of the inner region of the flow at time $t=10 t_a$ (bottom panel in Fig. \ref{fig:density_iso}). The arrows indicate 
the velocity vectors of the shocked matter inside the cocoon.  The velocity vectors of the 
unshocked wind were omitted for clarity.  An equatorial stream of dense matter towards the inner boundary is  clearly visible.}
\label{fig:Zoom}
\end{figure}

\subsection{Case A: Isothermal gas }
\label{sec:CaseA}
\begin{figure}
\includegraphics[width=8cm]{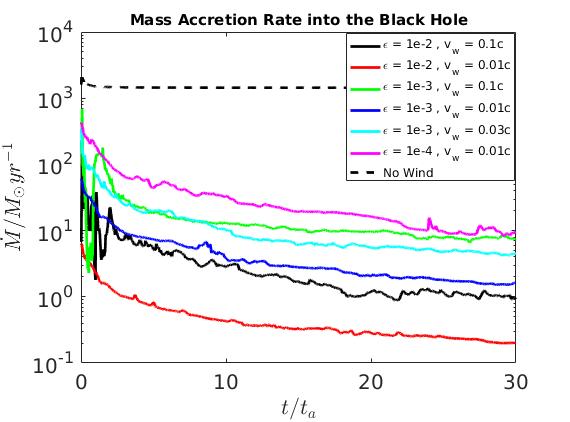}
\caption{Time evolution of the mass accretion rate $\dot{M}_{BH}$ (in absolute value) in case A when wind injection is switched off (dashed line) and on (solid lines).  The different colours correspond the the cases studied in table \ref{table:T1}}.
\label{fig:dotM_iso}
\end{figure}

Since radiative cooling is not included in this numerical experiment, the ambient gas was taken
to be cold initially to activate accretion. We find that the results are independent of
the initial pressure as long as $p(t=0)<<\rho \sigma^2$.  
Snapshots from a simulation with the fiducial values $\epsilon=10^{-2}$ and $v_w=0.1 c$,
each showing density (right half) and temperature (left half) maps, are displayed in Fig. \ref{fig:density_iso}
 (see Fig \ref{app:fig:scheme} for a schematic diagram of the different flow components).
A strong collimation of the unshocked wind is clearly seen, which is the reason for the elongated cocoon.
Such strong collimation is featured in all the cases we explored, and appears to be generic. 
The velocity of the contact discontinuity slightly changes with time due to the intermittent accretion in
the initial accretion burst, with an average value of $v_h\simeq 13 \sigma$ at time $t=10t_a$. 
As also seen from Fig \ref{fig:density_iso}, the two cocoons that inflate above and below the equatorial plane merge 
at early time, forming an equatorial bridge of dense matter  which is ultimately pulled in by the gravitational 
force, and gets accreted by the SMBH.   This inflow of shocked matter is evident in 
the enlarged view displayed in Fig \ref{fig:Zoom}, where velocity vectors are indicated by arrows.  
We find this accretion mode to be quite stable following the initial phase (Fig. \ref{fig:dotM_iso}).
 The mass accretion rate appears to be strongly suppressed by the wind feedback.  The black solid line in 
Fig. \ref{fig:dotM_iso} indicates that it is smaller by a factor
of $\chi\equiv \dot{M}_{max}/\dot{M}_{BH} \simeq 10^3$ than the value obtained when wind injection is switched off (Eq. (\ref{eq:dotM_iso})), 
consistent with the value derived in LN18 for the same parameters.    Note that the actual suppression, of  $\dot{M}_{in}$, is a factor of 3 smaller
for this choice of parameters.  The other lines in Fig. \ref{fig:dotM_iso} 
correspond to the different cases listed in table \ref{table:T1}, as indicated in the figure legend.

It is instructive to compare the velocity of the wind's head with the analytic result derived in appendix \ref{app:wind}.
The ratio of the average wind and ambient gas densities can be computed in terms 
of the ratio $\kappa =\dot{M}_{in}/\dot{M}_{max}$ measured  in the simulation. 
The average mass flux of the wind at radius $r$ can be expressed as $\dot{M}_w = \rho_w v_w\pi a^2$, where $a(r)$ is the cross
sectional radius of the wind at $r$.  Combined with
Eqs. (\ref{eq:dotM_in}) and (\ref{eq:dotM_iso}) one finds:
\begin{equation}
\rho_w/\rho_g \simeq \frac{2 \epsilon}{2\epsilon+(v_w/c)^2}\left(\frac{\sigma}{v_w}\right)\frac{2\kappa}{(a/r)^2}.
\label{eq:rho_w/rho_g}
\end{equation}
 For our choice of fiducial parameters we find $\kappa\simeq 2.5\times10^{-3}$ and $a/r=0.05$ at time $t=30t_a$,  
which yields $\rho_w/\rho_g\simeq 1.3\sigma/v_w$.  
The head velocity is given to a good approximation by 
$v_h=v_w/(1+\sqrt{\rho_g/\rho_w})\simeq \sqrt{\rho_w/\rho_g} v_w$ (see appendix \ref{app:wind} for details).
Thus,  $v_h\simeq\sqrt{1.3\sigma v_w} \simeq 11.4 \sigma$, in good agreement with the measured value ($12.5\sigma$).

\begin{figure}
\includegraphics[width=8cm]{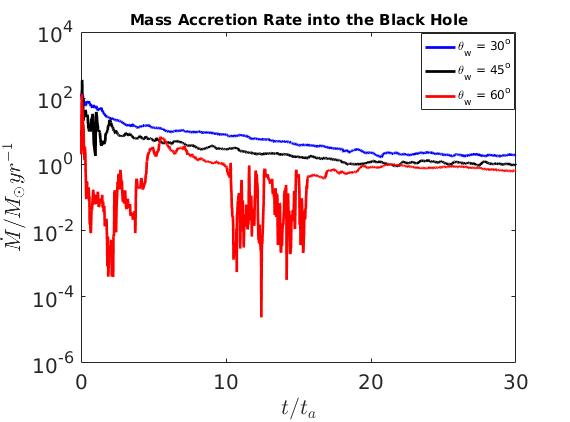}
\caption{Same as Fig. \ref{fig:dotM_iso} for the fiducial simulation, but with different values of the wind opening angle $\theta_w$, as indicated.}
\label{fig:dotM_angles}
\end{figure}	


\begin{figure*}
\includegraphics[width=6cm]{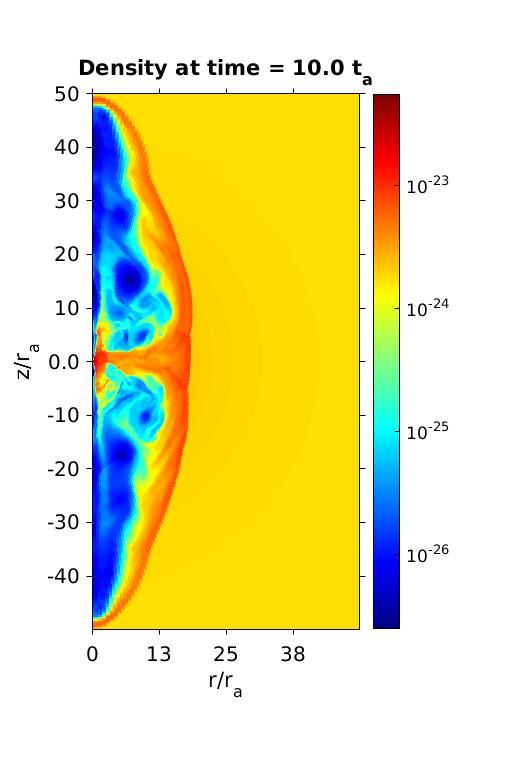}  \includegraphics[width=6cm]{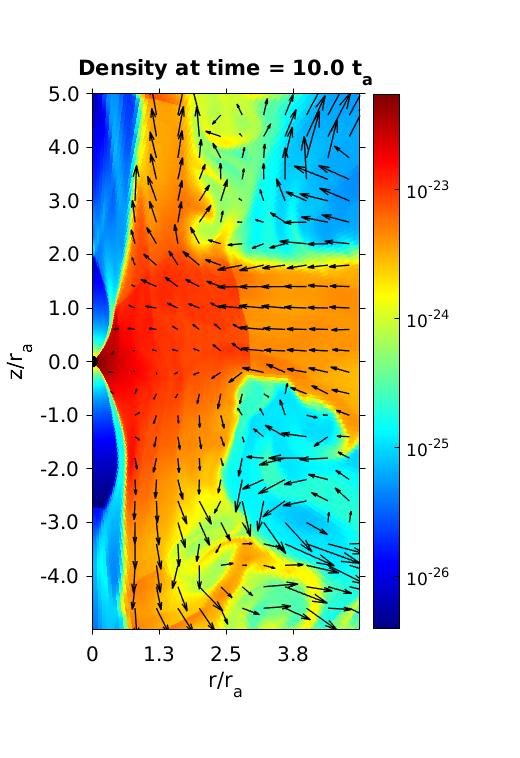}  
\caption{Left: Density map at $t=10t_a$ from the fiducial simulation in case B.   Right: Enlarged view of the inner
region of the flow, with superposed velocity vectors (omitted in the wind sector for clarity) }
\label{fig:density_unif}
\end{figure*}

\begin{figure}
\includegraphics[width=8cm]{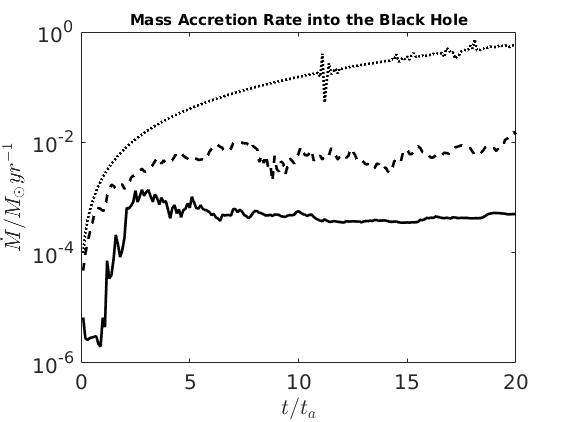}
\caption{Time evolution of mass accretion rate $\dot{M}_{BH}$ (in absolute value) in case B, for $\epsilon=10^{-2}, v_w=0.1c$ (solid line)
and $\epsilon=10^{-4}, v_w=0.01c$ (dashed line).  The dotted line depicts the evolution when 
wind injection is switched off. }
\label{fig:dotM_unif}
\end{figure}

To study the dependence of the accretion rate on wind properties we have run simulations with different 
values of $\epsilon$ and $v_w$.   Those encompass parameters typical to BAL QSO winds  \citep[e.g.,][]{borguet2013,chamberlain2015,williams2016} 
and ultra-fast outflows \citep[e.g.,][]{pounds2009,tombesi2010,maiolino2012,tombesi2015,bischetti2018}. 
The results are summarized in table \ref{table:T1}, and compared with 
the analytic result derived in LN18.   The corresponding Eddington ratios,  $\dot{m}_{BH} =\dot{M}_{BH}/\dot{M}_{Edd}$, 
here $\dot{M}_{Edd} = 2.3\,  M_\odot$ yr$^{-1}$ for $M_8=1$ and an assumed radiative 
efficiency of $0.1$, are also listed.   It is worth noting that cases with $\dot{M}_{BH}/\dot{M}_w >>1$ may not be realistic,
but they are, nonetheless,  included in the table as case study.  
We find a good agreement with the analytic results derived in LN18 for the realistic cases, $\dot{M}_{BH}/\dot{M}_{w}<1$. In particular, for a fixed
value of $\dot{M}_{BH}/\dot{M}_{w}$ the accretion rate at the inner boundary (as well as onto the SMBH) scales roughly as $\sqrt{\epsilon}$. For
the cases with $\dot{M}_{BH}/\dot{M}_{w}> 1$ we find the same trend as in LN18, but with overall lower accretion rates. 
Note that one can formally write $\dot{M}_{BH}=\dot{M}_{in}/(1+\dot{M}_w/\dot{M}_{BH})$.  Thus, in the regime $\dot{M}_{BH}/\dot{M}_{w} \gg 1$
the injected wind power, $L_w=\epsilon\dot{M}_{BH}c^2$, is proportional to $\dot{M}_{in}$, whereas in the regime $\dot{M}_{BH}/\dot{M}_{w}\ll 1$ the
wind power depends also on the ratio $\dot{M}_{BH}/\dot{M}_{w}$.   We attribute the somewhat different scaling
of accretion rate with wind parameters in these two regimes to this effect. 
The dependence of accretion rate on wind parameters observed in our simulations is somewhat different than that reported 
by \cite{ostriker2010} for a spherical wind.
When the wind becomes too weak, such that the shock velocity does not exceed $\sigma$ significantly, the accretion rate starts approaching $\dot{M}_{max}$.    We find this transition to be quite abrupt;  for $v_w =0.1c$,  $\epsilon\simlt10^{-4}$ the shock velocity is around $1\sigma$, and the wind injection process becomes highly intermittent, switching
on and off sporadically.  At $\epsilon=4\times10^{-5}$ the accretion rate exceeds $200~M_\odot$ yr$^{-1}$ (ten times larger than for $\epsilon=10^{-4}$, 
while  for $\epsilon\simeq3\times 10^{-5}$ the wind is completely suffocated and there is no suppression at all ($\dot{M}_{in}=\dot{M}_{max}$).    A similar
behaviour was reported by \cite{costa2014}.

 Finally, we examined the dependence of the accretion rate on the opening angle of the wind.  The accretion profiles obtained for
the fiducial parameters and different values of $\theta_w$ are exhibited in Fig. \ref{fig:dotM_angles}.   The asymptotic values are
$2, 1.1$ and $0.81~ M_\odot$ yr$^{-1}$ for $\theta_w=30^\circ, 45^\circ$ and $60^\circ$, respectively. 
It indicates that the values of $\dot{M}_{BH}$ in the relaxed state are insensitive to $\theta_w$.

\begin{table*}
\begin{center}\label{table:T1}
Case A simulations: Isothermal gas without cooling\\
\begin{tabular}{lllllllll}
\hline\hline
$\epsilon$ & $v_w/c$ & $\dot{M}_{in}(M_\odot/yr)$ &$\dot{M}_{BH}$ ($M_\odot/yr$) & $\dot{M}_{BH}/\dot{M}_{Edd}$ &$\dot{M}_{BH}$ ($M_\odot/yr$) & $\dot{M}_{BH}/\dot{M}_{w}$ &  $\dot{M}_{BH}/\dot{M}_{max}$ & $\dot{M}_{in}/\dot{M}_{max}$\\ 
 &  & Simulation & Simulation & Simulation & LN18 & Simulation & Simulation & Simulation\\
\hline
$10^{-2}$ & $0.1 $ &3.3 & 1.1 & 0.48 & 3.8 &0.5 & 0.0009 & 0.0025\\
$10^{-3}$ & $0.1 $ &  9.8 & 8.2& 3.6 & 38 & 5& 0.006 & 0.007\\
$10^{-4}$ & $0.1 $ & 20.4 & 20& 8.7 & 385 & 50& 0.015 & 0.015\\
$10^{-3}$ & $0.03 $ & 16 & 5  & 2.2 & 11.4 & 0.5 & 0.004 &0.012\\
$10^{-2}$ & $0.01 $ & 48.2 & 0.24 &  0.1& 0.38 &0.005 & 0.0002 & 0.038\\
$10^{-3}$ & $0.01 $ & 38 & 1.8 & 0.78& 3.8 & 0.05 & 0.0015 & 0.03\\
$10^{-4}$ & $0.01 $ &33 & 11& 4.8& 38 &0.5 & 0.009 & 0.025\\
\hline
\end{tabular}
\caption{Summary of the simulation results for case A.  For a fixed $\dot{M}_{BH}/\dot{M}_{w}$ value the accretion rate onto the inner boundary $\dot{M}_{in}$ scales roughly as $\sqrt{\epsilon}$.  Note that $\dot{M}_{w}/\dot{M}_{BH}=2\epsilon c^2/v_w^2$.
}
\end{center}
\end{table*}

\subsection{Case B: uniform initial state}
\label{sec:caseB}
As in case A, the gas is taken to be initially at rest and cold ($kT<< m_p\sigma^2$).  In our fiducial simulation
the initial gas density is $\rho_0/m_p=1$ cm$^{-3}$, $\epsilon=10^{-2}$ and $v_w=0.1 c$.
Since for a conical wind the density ratio $\rho_w/\rho_g \propto r^{-2}$, it is naively expected that the wind 
will undergo a strong collimation.  Indeed, we find that this occurs  already at early stages, as seen in Fig \ref{fig:density_unif}.  
This gives rise to the highly elongated cocoon seen in the figure, and to a nearly constant head velocity of $4.5\sigma$.   
The temporal evolution of the mass accretion rate is shown as a solid line in Fig. \ref{fig:dotM_unif}, where it is compared with the 
accretion profile in the absence of a wind (dotted line).   As seen, in the presence of feedback, the accretion rate approaches 
a constant value of $6\times10^{-4} M_\odot$ yr$^{-1}$ after a few $t_a$.   Note that for this choice of parameters $\dot{M}_{in}=3\dot{M}_{BH}$.
The saturation of the accretion rate implies a suppression that grows with time roughly as $t^2$  and can reach huge values on relatively short time scales. For example after $t\simeq20t_a = 0.4$ Myr the suppression is already by a factor of $10^3$ compared with the maximal possible accretion rate.
The prime reason is that the expansion of the cocoon precedes  that of the accretion front, implying that the density ahead of the forward shock 
does not have time to grow significantly beyond its initial value.   Consequently, the accretion rate is a fraction of the rate 
$\dot{M}_0=4\pi\rho_0\sigma r_a^2$,  which is constant in time.  
This should be compared to the maximal accretion rate obtained in the absence of wind feedback, Eq. (\ref{app:eq:Mdot_ev}), that 
evolves as $\dot{M}_0 (t/t_a)^2$ with time.
The terminal value of $\dot{M}_{BH}$ depends on wind parameters, but only moderately.
The dashed line in Fig.  \ref{fig:dotM_unif} delineates the result of a run with $\epsilon=10^{-4}$, $v_w=10^{-2}c$. As seen, it features a very similar 
accretion profile, with a terminal value larger by a factor of about 7 than the fiducial run, consistent with the result of \cite{negri2017}.  These values are smaller than the accretion rates inferred for 
high redshift AGNs (at $z>2$, see Fig. 1 in LN18) by several orders of magnitudes. 

\subsection{Scaling of the simulation}
\label{sec:scaling}
From the above parametrization one readily obtains the scaling of the simulation results with $\sigma$. 
Upon normalizing velocities by $\sigma$ ($\tilde{v}=v/\sigma$), radii by $r_a$, densities by $\rho_0$, accretion rates by $\dot{M}_0=4\pi\rho_0\sigma r_a^2$,
and power by $L_0=4\pi \rho_0\sigma^3 r_a^2$, the relations $\dot{m}_{BH}= \dot{m}_{in}/(1+2\epsilon c^2/\sigma^2\tilde{v}_w^2)$
and $l_w=(\epsilon c^2/\sigma^2)\dot{m}_{BH}$ are obtained, where $\dot{m}_{BH}=\dot{M}_{BH}/\dot{M}_0$, $\dot{m}_{in}=\dot{M}_{in}/\dot{M}_0$,
 and $l_w=L_w/L_0$. 
It is seen that the parameter $\sigma$ can be eliminated upon
redefining the efficiency factor according to $\tilde{\epsilon} = \epsilon/(\sigma/c)^2$, 
yielding a universal model that depends only on the normalized wind parameters,
$\tilde{\epsilon}$ and $\tilde{v}_w$, and in particular  is independent of the velocity dispersion of the galaxy.   Note that in case A a convenient choice 
for the fiducial density is $\rho_0=f_g \sigma^2/2\pi G r_a^2$, indicating that $\dot{M}_0$ is the maximum rate $\dot{M}_{max}$ given by Eq. (\ref{eq:dotM_iso}),
which is independent of the SMBH mass.  Since the scalings of $\epsilon$ and $v_w$ depend solely on $\sigma$, it implies that $\dot{M}_{in}$
and $\dot{M}_{BH}$ are independent of $M_{BH}$ as well in case A.  This is no longer true in case $B$, where the initial density introduces another scale into the problem.   Note also that if the scaling $\dot{M}_{in}\propto\epsilon^q$ is found for fixed values of $\sigma$ and 
$\dot{M}_{BH}/\dot{M}_{w}$, it can be translated into the scaling $\dot{M}_{in}\propto\sigma^{3+2q}$ at fixed 
values of $\epsilon$ and $v_w$.

\begin{figure}
\includegraphics[width=8cm]{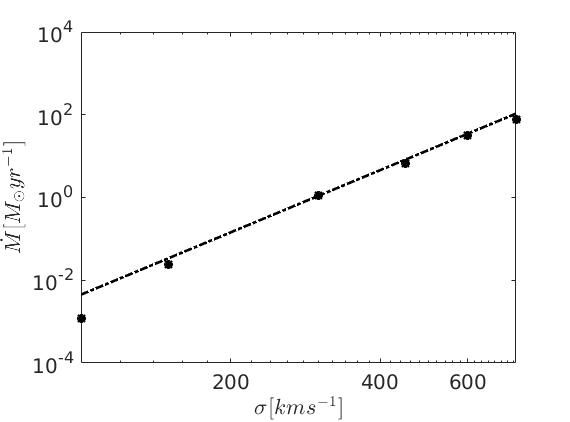}
\caption{Dependence of SMBH accretion rate, $\dot{M}_{BH}$, on velocity dispersion $\sigma$ (black circles). The
dotted line marks the relation $\dot{M} \propto \sigma^5$, and is included to guide the eye. }
\label{fig:dotM_sigma}
\end{figure}

Applying this scaling to case A implies that our finding that 
for a fixed  $\dot{M}_{BH}/\dot{M}_{w}$ value the accretion rate scales roughly as $\sqrt{\epsilon}$ means that
in case A the dependence of $\dot{M}_{BH}$ on $\sigma$ should be steeper than $\sigma^3$. To check this,
we repeated the fiducial simulation ($\epsilon=10^{-2}, v_w=0.1c$) with  different values of $\sigma$. 
The result, exhibited in Fig \ref{fig:dotM_sigma}, indicates that $\dot{M}_{BH}\propto\sigma^5$.   This dependence is
slightly steeper than expected from the table, however, note that the table doesn't cover the values that correspond to 
$\sigma=100$ and $150$.  Indeed, if these two points are excluded the fit is closer to $\sigma^4$.  Note also that the time 
required to reach a quasi-steady state (complete decay of the initial  transient) scales as $\sigma^{-2}$, hence much longer runtimes
are needed for the low sigma runs to reach the final values, and it could be that the result is somewhat affected by this.  
The latter scaling breaks down in the supercritical regime ($\sigma_{300}>1$)
if the accretion into the SMBH is capped at the Eddington limit.
 
The above results imply that  for an isothermal gas distribution, feedback introduces a robust suppression of
gas inflow within the expanding cocoon.  The amount that will ultimately be absorbed by the SMBH 
depends primarily on the physics of the accretion flow in the vicinity of the black hole. 

\subsection{Effect of resolution}
To examine the effect of resolution on the evolution of the system we made runs with varying number of grid points. 
 We have performed several different tests, in some keeping the same number of angular cells and changing the radial 
grid and in others vice versa.  In those tests the resolution was increased until we reached convergence.
We find that reducing the resolution of the radial grid (keeping the same angular grid) merely leads (except for the expected loss of structure) to
a modest increase of the accretion rate.  Our convergence test indicates a reduction of about $15\%$ in the accretion rate
in the fiducial simulations when the resolution of the uniform patch was increased from 250 to 1000 gridpoints. 

\begin{figure}
\includegraphics[width=6cm]{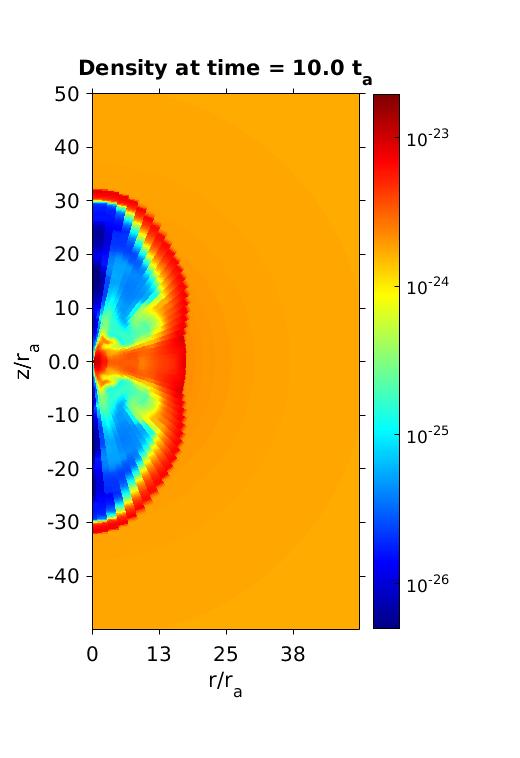}%
\caption{Same run as in Fig \ref{fig:density_unif} but with a $\theta$ setup of 50 gridpoints instead of $200$.  The 
lack of collimation due to the reduced resolution is evident.}
\label{fig:resol_test}
\end{figure}

More significantly, we find that  insufficient angular grid resolution near the axis results in less collimation of the wind
and a considerably slower shock velocity.  An example is shown in Fig. \ref{fig:resol_test}, where a run identical to 
the case shown in Fig. \ref{fig:density_unif} but with 50 angular gridpoints rather than 200 (and the same radial grid) is exhibited.   
The differences are apparent; the low resolution run features a 
round cocoon and a decelerating shock, whereas in the high resolution case the cocoon is elongated and the
shock velocity is roughly constant. 
The terminal accretion rate in the low resolution case is smaller by a factor of about 3 
compared with the high resolution case.   The increase in accretion rate with increasing resolution is a consequence of the
stronger collimation, that leads to a smaller energy deposition in the cocoon (as is evident from the difference in velocity of the 
wind's head between the high and low resolution runs (see Figs    \ref{fig:density_unif} and   \ref{fig:resol_test}).

 \subsection{Effect of cooling}
\label{sec:cooling}
In order to explore the effect of cooling on the evolution of the accretion rate, the fiducial simulation in Case A was repeated 
with cooling included.    In difference from the no-cooling runs, the gas in the bulge is taken initially to be in a hydrostatic equilibrium.
Accretion commences after a few cooling times during which the temperature near the sphere of influence is sufficiently reduced.
To simplify the numerics, we find it sufficient to incorporate only
bremsstrahlung cooling, which for the high gas densities  invoked in case A, is much faster than the wind 
propagation (see Eq. (\ref{eq:t_ff_iso})).  We used the power law cooling module implemented in PLUTO, and set the temperature 
floor at $5\times 10^5$ K.   We have made two runs, one with the resolution used in the cases with no cooling, and another one with
a resolution 4 times higher (400 gridpoints in $\theta$ and 2000 in the uniform patch of the radial grid).  We find some differences
in structure between the two runs, and a higher mean accretion rate in the low resolution case, 
but the overall trend is quite similar.   We think that the higher accretion rate in the low resolution case results from artificial 
stability of the dense filaments, that prevents their destruction.  As the resolution increases more clouds are prone to instabilities 
and ultimately crash.  Unfortunately, this very high resolution increases the computing time dramatically, and so 
we were able to run the simulation only up to a time of about $3t_a$. 
Below we present the results of the high resolution run.

Cooling of the shocked wind gas via inverse Compton scattering of the quasar radiation is ignored in the present study.
Weather it is important  \citep{king2003,king2015} or not  \citep{bourne2013} is yet an open issue.  From Eq. (\ref{app:eq:comp_cooling}) we estimate 
$v_w t_c/r\sim 0.05 \tilde{r}$, assuming strong coupling between electrons and ions.  This implies effective 
wind cooling up to $r\sim 20 r_a$ for our setup.     The latter estimate should be reduced by some factor if the equilibration time 
of ions is not much shorter than the wind expansion time \citep{faucher2012}.  On the other hand, beaming of the quasar radiation, that might 
enhance the luminosity in the polar region, can lead to a more effective cooling.   Based on these estimates we naively anticipate 
that wind cooling might only be important in the early stages of evolution, unless strong beaming ensue.   We intend to
add this to our model in a future work.

\begin{figure*}
\includegraphics[width=8cm]{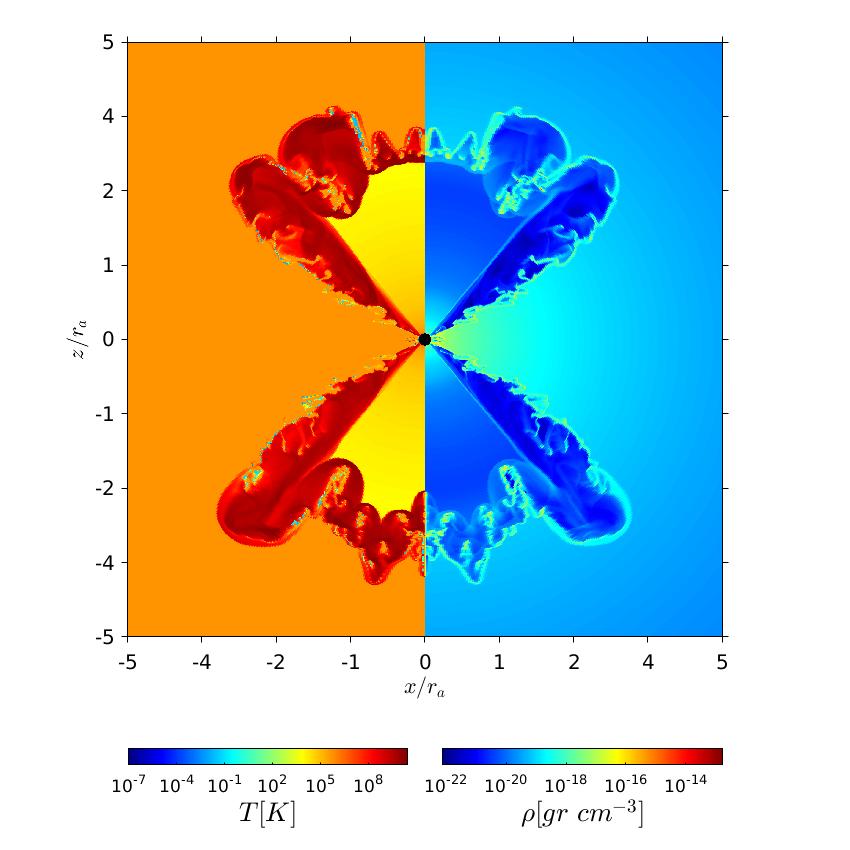}  \includegraphics[width=8cm]{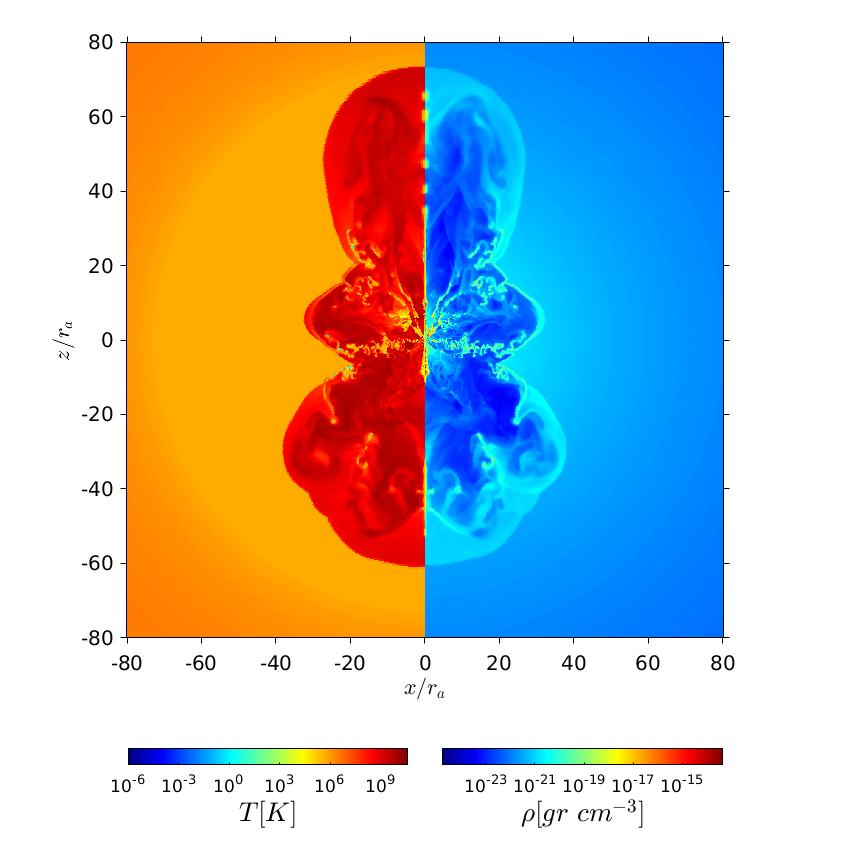} 
\caption{Density and temperature maps at times $t=0.2 t_a$ (left panel) and $2.5 t_a$ (right panel), 
for case A with bremsstrahlung cooling.  The yellowish
stripe that envelopes the  wind in the density map  in the left panel is the compressed, shocked ambient shell.  The fingers indicate an early onset of
the Rayleigh-Taylor instability at the contact surface, that leads to mixing of the shocked wind and shocked ambient material, 
as seen in the right panel (and more clearly in Fig. \ref{fig:ISO_cooling-enlarged}).
It also leads to deflection of streamlines of the unshocked wind, and its disruption.}
\label{fig:ISO_cooling-1}
\end{figure*}

\begin{figure}
\includegraphics[width=8cm]{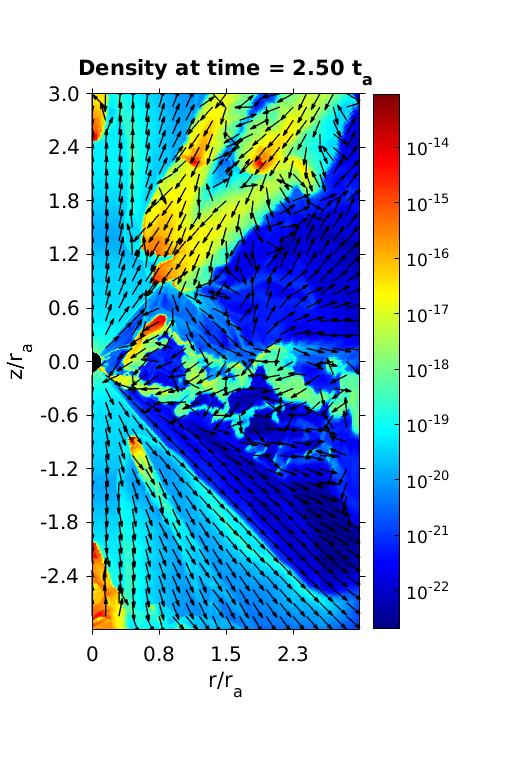}  
\caption{Enlarged view of the inner region of the flow at time $t=2.5t_a$.  
The inwards equatorial  stream of inhomogeneous matter seen in this plot gives rise to
intermittent accretion (Fig. \ref{fig:ISO_cooling-Mdot}).  The cold dense blobs seen at higher inclination angles 
move inwards rather slowly, at a velocity $\ll \sigma$.  These blobs may be sites of star formation.}
\label{fig:ISO_cooling-enlarged}
\end{figure}

Figure (\ref{fig:ISO_cooling-1}) displays density and temperature maps at times $t=0.2 t_a$ and $2.5 t_a$.  As seen, the shocked 
ambient gas is compressed to a very thin shell early on by virtue of the fast cooling.  This leads to a rapid growth of the 
Rayleigh-Taylor instability, as
clearly seen in the images in the two left panels.  Complete mixing of the shocked wind and shocked ambient matter is observed at 
later times (the cold dense blobs seen in the right panels).   This is seen more clearly in the enlarged view of the inner region
exhibited in Fig. \ref{fig:ISO_cooling-enlarged}.    We speculate that the cold, dense blobs seen in the plot 
may be associated with star-forming sites (see also \cite{nayakshin2012,zubovas2013,mukherjee2018}).   
However, it could well be that our resolution is insufficient to allow growth of local instabilities that
might destroy the clouds.  Such instabilities are expected to be generated at the interface of dense clouds by the engulfed wind. 
A rough estimate of the shock crossing time of a cloud of size $R_{cloud}$ and density $\rho_{cloud}$ is $t_{cross}\sim (\rho_{cloud}/\rho_w)^{1/2}(R_{cloud}/v_w) = (\rho_{cloud}/\rho_w)^{1/2}(R_{cloud}/r_a)(\sigma/v_w) t_a$.  For the wind velocity adopted in this 
example, $v_w/\sigma =100$, and cloud size $R_{cloud}/r_a <1$, even clouds of density $\rho_{cloud} \simgt 10^4\rho_w$ 
are expected to be shredded over time of a few $t_a$.   While we observed disruption of some blobs, most seem to survive though
the runtime.   As their mean velocity is $\ll \sigma$ due to the drag exerted by the engulfed wind, their ultimate fate is uncertain. 

\begin{figure}
\includegraphics[width=8cm]{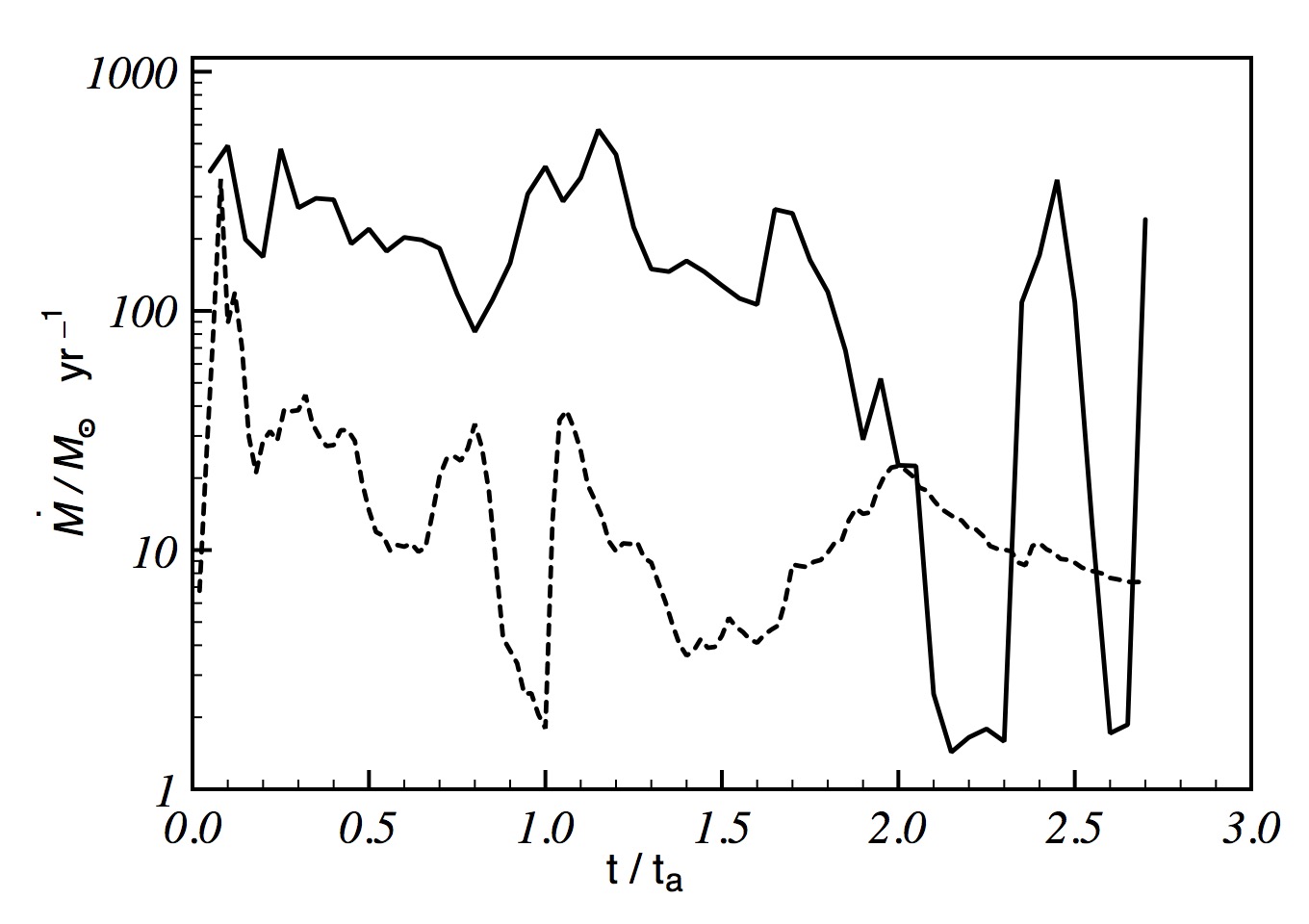}  
\caption{Temporal evolution of accretion rate in the presence of strong cooling (solid line). The dashed line corresponds to the fiducial
run with no cooling (the solid black line in Fig. \ref{fig:dotM_iso}), and is shown here for a comparison. }
\label{fig:ISO_cooling-Mdot}
\end{figure}

The temporal evolution of the accretion rate is shown in Fig. \ref{fig:ISO_cooling-Mdot} (solid line) and is compared with the 
no-cooling run for the same wind parameters.    As expected, it is highly intermittent
by virtue of the  strong inhomogeneity of the inflowing matter within the cocoon.  As in the no cooling case, we find that accretion 
onto the inner boundary occurs predominantly along an equatorial belt (Fig. \ref{fig:ISO_cooling-enlarged}).   The mean rate is initially high,
owing to the strong compression of the shocked ambient gas, that leads to a late 
merger of the two cocoons  compared with the no cooling case, but then declines over time as the cocoons merge and expand. 
The mean rate seems highly suppressed, as in the no cooling case, but given the limited runtime the actual value
is uncertain.   A more comprehensive analysis is left for a future work. 

 We note that, in practice, thermal instabilities may develop in the unshocked medium \citep{ciotti1997},
leading to formation a clumpy structure outside the cocoon that might affect its dynamics and the resultant 
accretion rate \citep[e.g.,][]{nayakshin2012,mukherjee2016}. This, however, requires the instability growth time to be short compared with the expansion time of the cocoon, and
the clumps to survive their interaction with the wind.  
The lack of heating processes in our simulation prevents such an occurrence, and further analysis is beyond the scope of this paper.

\section{Conclusions}
We conducted a numerical study of the effect of AGN wind feedback on the accretion onto the SMBH, that resolves scales much smaller than 
the SMBH sphere of influence ($\sim 5$ pc for a $10^8~M_\odot$ BH). We considered two initial gas density profiles; uniform and isothermal. We studied first, in details, accretion when cooling of the  gas is ignored, and then considered the effect of bremsstrahlung cooling on the feedback process.  The latter analysis should be considered preliminary, as it is  limited by the relatively short runtime and the neglect of Compton
scattering of the quasar emission by the shocked wind plasma.

The main conclusion to be drawn from our analysis is that strong suppression of the mass accretion rate by feedback is 
generally anticipated, despite the strong collimation of the wind observed in essentially all of the cases explored.  The level of suppression depends on the density profile of the accreted gas.  For initially uniform gas density, $\rho_0$, the mass accretion rate was found to be constant in time and of the order of $\dot{M}_0=4\pi \rho_0 \sigma r_a^2$, compared to the accretion with no wind that grows as $t^2$. For typical ISM densities, $\rho_0/m_p\sim 1 {\rm~cm^{-3}}$, this rate is smaller by several orders of magnitudes than the typical accretion rates  observed in samples of high redshift AGNs (see LN18, and references therein). 

 When the initial gas density has an isothermal sphere profile, the maximal accretion rate (in the absence of feedback) is the spherical free-infall rate (Eq. (\ref{eq:dotM_iso})). We have found that over a broad range of wind parameters (i.e., $\epsilon$ and $v_w$), that encompass values inferred 
from observations of BAL QSO winds and ultra-fast outflows, wind feedback suppresses the net mass infall rate to less than $\sim 3\%$ of the 
maximal rate.  This suggests that subgrid prescriptions for the accretion rate in large scale cosmological simulations should not exceed this value (although cooling can somewhat alter this value). 
The regulated mass infall rate is independent of the SMBH mass, scales roughly as $\sigma^5$, and its dependence 
on the wind parameters is consistent with that expected from momentum balance, as derived in LN18.
The actual mass absorption rate by the SMBH is determined by the accretion disk physics, which is beyond the scope of this paper. When using realistic parameters (see the cases in table \ref{table:T1} where $\dot{M}_{BH}/\dot{M}_w < 1$), the terminal SMBH accretion rate is between about $0.2$ and $20~ M_\odot$ yr$^{-1}$ for $\sigma=300 {\rm ~km/s}$. These rates are somewhat lower than those inferred from observations when cooling is ignored.   
Our preliminary study seems to indicate that the accretion rate increases by a factor of a few in the presence of rapid cooling, 
in better agreement with the observations. 
The corresponding Eddington ratios in table \ref{table:T1} span the range $0.1$ to about $10$
for the fiducial black hole mass adopted in the simulations ($M_{BH}=10^8~ M_\odot$). The possibility of super-Eddington accretion in 
these situations is discussed in \cite[e.g.,][]{volonteri2015,begelman2017,levinson2018}.  For larger  SMBHs ($M_{BH}>10^9$)
these accretion rates correspond to mildly sub-Eddington accretion, in accord with the Eddington ratios measured for high redshift ($2<z<7$)
AGNs \citep[][and references therein]{levinson2018}. 

The final SMBH mass will be limited by the net mass accreted over the time it takes the shock to cross the bulge.
The time it takes the shock to cross the bulge is $t_s \sim GM_b/v_s\sigma^2\sim  200 (\sigma/v_s)M_{b,12}\sigma_{300}^{-3}$ Myr,
where $M_b=10^{12} M_{b,12} M_\odot$ is the bulge mass, $v_s$ is the shock velocity, and typically $v_s/\sigma \simeq $ a few.  
The SMBH mass increment over this time is $\Delta M_{BH} \simeq \dot{M}_{BH} t_s \sim 10^{9} (\sigma/v_s) \sigma_{300}^2M_{b,12}~M_\odot$, adopting 
the scaling delineated in Fig. \ref{fig:dotM_sigma}, with $\dot{M}_{BH}=5 M_\odot$ yr$^{-1}$ at $\sigma_{300}=1$ (corresponding to a typical
BAL wind with $\epsilon=10^{-3}, v_w=0.03 c$, see table \ref{table:T1}).  With $v_s/\sigma \sim $ a few from the simulations,
this is somewhat smaller than the largest SMBH masses inferred from observations, or, when using the Faber-Jackson relation, 
from the Magorrian relation \citep[e.g.,][]{magorrian1998,kormendy2013}.   Again, rapid cooling is likely to 
give rise to a larger $\dot{M}_{BH}$ and a larger $\Delta M_{BH}$.  Furthermore, the above estimate applies to a single episode,
while it could well be that the SMBH growth occurs over several merger episodes, as discussed in LN18.  Observationally, the flat accretion phase
is seen only at redshifts $z>2$, below which the mean rate rapidly declines (see data in LN18). We interpret this as indicating enhanced 
cosmic (merger) activity at early epochs, during which growth occurs.

Inclusion of cooling leads to a rapid growth of the Rayleigh-Taylor instability at the contact interface early on, followed by complete
mixing of the shocked wind and ambient gas.   The strong inhomogeneity of the inflowing matter gives rise to a highly intermittent accretion, 
however, we find that the mean rate remains highly suppressed.  At the end of our simulation the average accretion rate is higher by a factor of 
about 10 than in the no cooling case.  Given our limited runtime, and the lack of Compton cooling and heating in addition to bremsstrahlung losses, farther study is required to confirm these preliminary results.

 The above estimates ignore the effect of self-gravity \citep[e.g.,][]{pringle1981,lodato2007}, that may give 
rise to fragmentation of the disk at large radii, thereby
reducing the mass accretion rate.  However, this may only be relevant for extremely thin disks ($H/r\ll 1$).  For instance, at a radius
of $r\sim10^3 r_g$ from a $10^8~M_\odot$ SMBH, the Toomre criterion is satisfied provided $H/r<10^{-3}$ \citep{lodato2007}.  Wether such conditions 
can prevail in those systems, particularly given the heating expected by the quasar emission, is highly questionable.

\section*{Acknowledgements}
 AL thanks Hamid Hamidani for enlightening discussions and help.  
We also wish to thank Yishay Vadai for technical help,
and the anonymous referee for insightful and helpful comments.
Support by The Israel Science Foundation (grant 1114/17) is acknowledged. 

\bibliographystyle{mnras}

\begin{thebibliography}{}
\makeatletter
\relax
\def\mn@urlcharsother{\let\do\@makeother \do\$\do\&\do\#\do\^\do\_\do\%\do\~}
\def\mn@doi{\begingroup\mn@urlcharsother \@ifnextchar [ {\mn@doi@}
  {\mn@doi@[]}}
\def\mn@doi@[#1]#2{\def\@tempa{#1}\ifx\@tempa\@empty \href
  {http://dx.doi.org/#2} {doi:#2}\else \href {http://dx.doi.org/#2} {#1}\fi
  \endgroup}
\def\mn@eprint#1#2{\mn@eprint@#1:#2::\@nil}
\def\mn@eprint@arXiv#1{\href {http://arxiv.org/abs/#1} {{\tt arXiv:#1}}}
\def\mn@eprint@dblp#1{\href {http://dblp.uni-trier.de/rec/bibtex/#1.xml}
  {dblp:#1}}
\def\mn@eprint@#1:#2:#3:#4\@nil{\def\@tempa {#1}\def\@tempb {#2}\def\@tempc
  {#3}\ifx \@tempc \@empty \let \@tempc \@tempb \let \@tempb \@tempa \fi \ifx
  \@tempb \@empty \def\@tempb {arXiv}\fi \@ifundefined
  {mn@eprint@\@tempb}{\@tempb:\@tempc}{\expandafter \expandafter \csname
  mn@eprint@\@tempb\endcsname \expandafter{\@tempc}}}

\bibitem[\protect\citeauthoryear{{Begelman}}{{Begelman}}{2012}]{begelman2012}
{Begelman} M.~C.,  2012, \mn@doi [\mnras] {10.1111/j.1365-2966.2011.20071.x},
  \href {http://adsabs.harvard.edu/abs/2012MNRAS.420.2912B} {420, 2912}

\bibitem[\protect\citeauthoryear{{Begelman} \& {Volonteri}}{{Begelman} \&
  {Volonteri}}{2017}]{begelman2017}
{Begelman} M.~C.,  {Volonteri} M.,  2017, \mn@doi [\mnras]
  {10.1093/mnras/stw2446}, \href
  {http://adsabs.harvard.edu/abs/2017MNRAS.464.1102B} {464, 1102}

\bibitem[\protect\citeauthoryear{{Begelman}, {Blandford}  \& {Rees}}{{Begelman}
  et~al.}{1984}]{begelman1984}
{Begelman} M.~C.,  {Blandford} R.~D.,   {Rees} M.~J.,  1984, \mn@doi [Reviews
  of Modern Physics] {10.1103/RevModPhys.56.255}, \href
  {http://adsabs.harvard.edu/abs/1984RvMP...56..255B} {56, 255}

\bibitem[\protect\citeauthoryear{{Bischetti}, {Maiolino}, {Fiore}, {Piconcelli}
   \& {Fluetsch}}{{Bischetti} et~al.}{2018}]{bischetti2018}
{Bischetti} M.,  {Maiolino} R.,  {Fiore} S.~C.~F.,  {Piconcelli} E.,
  {Fluetsch} A.,  2018, preprint, \href
  {http://adsabs.harvard.edu/abs/2018arXiv180600786B} {} (\mn@eprint {arXiv}
  {1806.00786})

\bibitem[\protect\citeauthoryear{{Borguet}, {Arav}, {Edmonds}, {Chamberlain}
  \& {Benn}}{{Borguet} et~al.}{2013}]{borguet2013}
{Borguet} B.~C.~J.,  {Arav} N.,  {Edmonds} D.,  {Chamberlain} C.,   {Benn} C.,
  2013, \mn@doi [\apj] {10.1088/0004-637X/762/1/49}, \href
  {http://adsabs.harvard.edu/abs/2013ApJ...762...49B} {762, 49}

\bibitem[\protect\citeauthoryear{{Bourne} \& {Nayakshin}}{{Bourne} \&
  {Nayakshin}}{2013}]{bourne2013}
{Bourne} M.~A.,  {Nayakshin} S.,  2013, \mn@doi [\mnras]
  {10.1093/mnras/stt1739}, \href
  {http://adsabs.harvard.edu/abs/2013MNRAS.436.2346B} {436, 2346}

\bibitem[\protect\citeauthoryear{{Bourne}, {Nayakshin}  \& {Hobbs}}{{Bourne}
  et~al.}{2014}]{bourne2014}
{Bourne} M.~A.,  {Nayakshin} S.,   {Hobbs} A.,  2014, \mn@doi [\mnras]
  {10.1093/mnras/stu747}, \href
  {http://adsabs.harvard.edu/abs/2014MNRAS.441.3055B} {441, 3055}

\bibitem[\protect\citeauthoryear{{Bourne}, {Zubovas}  \& {Nayakshin}}{{Bourne}
  et~al.}{2015}]{bourne2015}
{Bourne} M.~A.,  {Zubovas} K.,   {Nayakshin} S.,  2015, \mn@doi [\mnras]
  {10.1093/mnras/stv1730}, \href
  {http://adsabs.harvard.edu/abs/2015MNRAS.453.1829B} {453, 1829}

\bibitem[\protect\citeauthoryear{{Chamberlain}, {Arav}  \&
  {Benn}}{{Chamberlain} et~al.}{2015}]{chamberlain2015}
{Chamberlain} C.,  {Arav} N.,   {Benn} C.,  2015, \mn@doi [\mnras]
  {10.1093/mnras/stv572}, \href
  {http://adsabs.harvard.edu/abs/2015MNRAS.450.1085C} {450, 1085}

\bibitem[\protect\citeauthoryear{{Choi}, {Ostriker}, {Naab}  \&
  {Johansson}}{{Choi} et~al.}{2012}]{choi2012}
{Choi} E.,  {Ostriker} J.~P.,  {Naab} T.,   {Johansson} P.~H.,  2012, \mn@doi
  [\apj] {10.1088/0004-637X/754/2/125}, \href
  {http://adsabs.harvard.edu/abs/2012ApJ...754..125C} {754, 125}

\bibitem[\protect\citeauthoryear{{Choi}, {Naab}, {Ostriker}, {Johansson}  \&
  {Moster}}{{Choi} et~al.}{2014}]{choi2014}
{Choi} E.,  {Naab} T.,  {Ostriker} J.~P.,  {Johansson} P.~H.,   {Moster} B.~P.,
   2014, \mn@doi [\mnras] {10.1093/mnras/stu874}, \href
  {http://adsabs.harvard.edu/abs/2014MNRAS.442..440C} {442, 440}

\bibitem[\protect\citeauthoryear{{Ciotti} \& {Ostriker}}{{Ciotti} \&
  {Ostriker}}{1997}]{ciotti1997}
{Ciotti} L.,  {Ostriker} J.~P.,  1997, \mn@doi [\apjl] {10.1086/310902}, \href
  {http://adsabs.harvard.edu/abs/1997ApJ...487L.105C} {487, L105}

\bibitem[\protect\citeauthoryear{{Ciotti}, {Pellegrini}, {Negri}  \&
  {Ostriker}}{{Ciotti} et~al.}{2017}]{ciotti2017}
{Ciotti} L.,  {Pellegrini} S.,  {Negri} A.,   {Ostriker} J.~P.,  2017, \mn@doi
  [\apj] {10.3847/1538-4357/835/1/15}, \href
  {http://adsabs.harvard.edu/abs/2017ApJ...835...15C} {835, 15}

\bibitem[\protect\citeauthoryear{{Costa}, {Sijacki}  \& {Haehnelt}}{{Costa}
  et~al.}{2014}]{costa2014}
{Costa} T.,  {Sijacki} D.,   {Haehnelt} M.~G.,  2014, \mn@doi [\mnras]
  {10.1093/mnras/stu1632}, \href
  {http://adsabs.harvard.edu/abs/2014MNRAS.444.2355C} {444, 2355}

\bibitem[\protect\citeauthoryear{{Debuhr}, {Quataert}  \& {Ma}}{{Debuhr}
  et~al.}{2011}]{debuhr2011}
{Debuhr} J.,  {Quataert} E.,   {Ma} C.-P.,  2011, \mn@doi [\mnras]
  {10.1111/j.1365-2966.2010.17992.x}, \href
  {http://adsabs.harvard.edu/abs/2011MNRAS.412.1341D} {412, 1341}

\bibitem[\protect\citeauthoryear{{Di Matteo}, {Springel}  \& {Hernquist}}{{Di
  Matteo} et~al.}{2005}]{diMatteo2005}
{Di Matteo} T.,  {Springel} V.,   {Hernquist} L.,  2005, \mn@doi [\nat]
  {10.1038/nature03335}, \href
  {http://adsabs.harvard.edu/abs/2005Natur.433..604D} {433, 604}

\bibitem[\protect\citeauthoryear{{Dubois}, {Peirani}, {Pichon}, {Devriendt},
  {Gavazzi}, {Welker}  \& {Volonteri}}{{Dubois} et~al.}{2016}]{dubois2016}
{Dubois} Y.,  {Peirani} S.,  {Pichon} C.,  {Devriendt} J.,  {Gavazzi} R.,
  {Welker} C.,   {Volonteri} M.,  2016, \mn@doi [\mnras]
  {10.1093/mnras/stw2265}, \href
  {http://adsabs.harvard.edu/abs/2016MNRAS.463.3948D} {463, 3948}

\bibitem[\protect\citeauthoryear{{Faucher-Gigu{\`e}re} \&
  {Quataert}}{{Faucher-Gigu{\`e}re} \& {Quataert}}{2012}]{faucher2012}
{Faucher-Gigu{\`e}re} C.-A.,  {Quataert} E.,  2012, \mn@doi [\mnras]
  {10.1111/j.1365-2966.2012.21512.x}, \href
  {http://adsabs.harvard.edu/abs/2012MNRAS.425..605F} {425, 605}

\bibitem[\protect\citeauthoryear{{King}}{{King}}{2003}]{king2003}
{King} A.,  2003, \mn@doi [\apjl] {10.1086/379143}, \href
  {http://adsabs.harvard.edu/abs/2003ApJ...596L..27K} {596, L27}

\bibitem[\protect\citeauthoryear{{King}}{{King}}{2010}]{king2010}
{King} A.~R.,  2010, \mn@doi [\mnras] {10.1111/j.1365-2966.2009.16013.x}, \href
  {http://adsabs.harvard.edu/abs/2010MNRAS.402.1516K} {402, 1516}

\bibitem[\protect\citeauthoryear{{King} \& {Pounds}}{{King} \&
  {Pounds}}{2015}]{king2015}
{King} A.,  {Pounds} K.,  2015, \mn@doi [\araa]
  {10.1146/annurev-astro-082214-122316}, \href
  {http://adsabs.harvard.edu/abs/2015ARA%26A..53..115K} {53, 115}

\bibitem[\protect\citeauthoryear{{Kormendy} \& {Ho}}{{Kormendy} \&
  {Ho}}{2013}]{kormendy2013}
{Kormendy} J.,  {Ho} L.~C.,  2013, \mn@doi [\araa]
  {10.1146/annurev-astro-082708-101811}, \href
  {http://adsabs.harvard.edu/abs/2013ARA%26A..51..511K} {51, 511}

\bibitem[\protect\citeauthoryear{{Kurk} et~al.,}{{Kurk}
  et~al.}{2007}]{kurk2007}
{Kurk} J.~D.,  et~al., 2007, \mn@doi [\apj] {10.1086/521596}, \href
  {http://adsabs.harvard.edu/abs/2007ApJ...669...32K} {669, 32}

\bibitem[\protect\citeauthoryear{{Levinson} \& {Nakar}}{{Levinson} \&
  {Nakar}}{2018}]{levinson2018}
{Levinson} A.,  {Nakar} E.,  2018, \mn@doi [\mnras] {10.1093/mnras/stx2542},
  \href {http://adsabs.harvard.edu/abs/2018MNRAS.473.2673L} {473, 2673}

\bibitem[\protect\citeauthoryear{{Lodato}}{{Lodato}}{2007}]{lodato2007}
{Lodato} G.,  2007, \mn@doi [Nuovo Cimento Rivista Serie]
  {10.1393/ncr/i2007-10022-x}, \href
  {http://adsabs.harvard.edu/abs/2007NCimR..30..293L} {30}

\bibitem[\protect\citeauthoryear{{Magorrian} et~al.,}{{Magorrian}
  et~al.}{1998}]{magorrian1998}
{Magorrian} J.,  et~al., 1998, \mn@doi [\aj] {10.1086/300353}, \href
  {http://adsabs.harvard.edu/abs/1998AJ....115.2285M} {115, 2285}

\bibitem[\protect\citeauthoryear{{Maiolino} et~al.,}{{Maiolino}
  et~al.}{2012}]{maiolino2012}
{Maiolino} R.,  et~al., 2012, \mn@doi [\mnras]
  {10.1111/j.1745-3933.2012.01303.x}, \href
  {http://adsabs.harvard.edu/abs/2012MNRAS.425L..66M} {425, L66}

\bibitem[\protect\citeauthoryear{{Mignone}, {Bodo}, {Massaglia}, {Matsakos},
  {Tesileanu}, {Zanni}  \& {Ferrari}}{{Mignone} et~al.}{2007}]{mignone2007}
{Mignone} A.,  {Bodo} G.,  {Massaglia} S.,  {Matsakos} T.,  {Tesileanu} O.,
  {Zanni} C.,   {Ferrari} A.,  2007, \mn@doi [\apjs] {10.1086/513316}, \href
  {http://adsabs.harvard.edu/abs/2007ApJS..170..228M} {170, 228}

\bibitem[\protect\citeauthoryear{{Mukherjee}, {Bicknell}, {Sutherland}  \&
  {Wagner}}{{Mukherjee} et~al.}{2016}]{mukherjee2016}
{Mukherjee} D.,  {Bicknell} G.~V.,  {Sutherland} R.,   {Wagner} A.,  2016,
  \mn@doi [\mnras] {10.1093/mnras/stw1368}, \href
  {http://adsabs.harvard.edu/abs/2016MNRAS.461..967M} {461, 967}

\bibitem[\protect\citeauthoryear{{Mukherjee}, {Bicknell}, {Wagner},
  {Sutherland}  \& {Silk}}{{Mukherjee} et~al.}{2018}]{mukherjee2018}
{Mukherjee} D.,  {Bicknell} G.~V.,  {Wagner} A.~Y.,  {Sutherland} R.~S.,
  {Silk} J.,  2018, \mn@doi [\mnras] {10.1093/mnras/sty1776}, \href
  {http://adsabs.harvard.edu/abs/2018MNRAS.479.5544M} {479, 5544}

\bibitem[\protect\citeauthoryear{{Nayakshin} \& {Zubovas}}{{Nayakshin} \&
  {Zubovas}}{2012}]{nayakshin2012}
{Nayakshin} S.,  {Zubovas} K.,  2012, \mn@doi [\mnras]
  {10.1111/j.1365-2966.2012.21950.x}, \href
  {http://adsabs.harvard.edu/abs/2012MNRAS.427..372N} {427, 372}

\bibitem[\protect\citeauthoryear{{Negri} \& {Volonteri}}{{Negri} \&
  {Volonteri}}{2017}]{negri2017}
{Negri} A.,  {Volonteri} M.,  2017, \mn@doi [\mnras] {10.1093/mnras/stx362},
  \href {http://adsabs.harvard.edu/abs/2017MNRAS.467.3475N} {467, 3475}

\bibitem[\protect\citeauthoryear{{Ostriker}, {Choi}, {Ciotti}, {Novak}  \&
  {Proga}}{{Ostriker} et~al.}{2010}]{ostriker2010}
{Ostriker} J.~P.,  {Choi} E.,  {Ciotti} L.,  {Novak} G.~S.,   {Proga} D.,
  2010, \mn@doi [\apj] {10.1088/0004-637X/722/1/642}, \href
  {http://adsabs.harvard.edu/abs/2010ApJ...722..642O} {722, 642}

\bibitem[\protect\citeauthoryear{{Pounds} \& {Reeves}}{{Pounds} \&
  {Reeves}}{2009}]{pounds2009}
{Pounds} K.~A.,  {Reeves} J.~N.,  2009, \mn@doi [\mnras]
  {10.1111/j.1365-2966.2009.14971.x}, \href
  {http://adsabs.harvard.edu/abs/2009MNRAS.397..249P} {397, 249}

\bibitem[\protect\citeauthoryear{{Pringle}}{{Pringle}}{1981}]{pringle1981}
{Pringle} J.~E.,  1981, \mn@doi [\araa] {10.1146/annurev.aa.19.090181.001033},
  \href {http://adsabs.harvard.edu/abs/1981ARA%26A..19..137P} {19, 137}

\bibitem[\protect\citeauthoryear{{Robertson}, {Bullock}, {Cox}, {Di Matteo},
  {Hernquist}, {Springel}  \& {Yoshida}}{{Robertson}
  et~al.}{2006}]{robertson2006}
{Robertson} B.,  {Bullock} J.~S.,  {Cox} T.~J.,  {Di Matteo} T.,  {Hernquist}
  L.,  {Springel} V.,   {Yoshida} N.,  2006, \mn@doi [\apj] {10.1086/504412},
  \href {http://adsabs.harvard.edu/abs/2006ApJ...645..986R} {645, 986}

\bibitem[\protect\citeauthoryear{{Schaye} et~al.,}{{Schaye}
  et~al.}{2015}]{schaye2015}
{Schaye} J.,  et~al., 2015, \mn@doi [\mnras] {10.1093/mnras/stu2058}, \href
  {http://adsabs.harvard.edu/abs/2015MNRAS.446..521S} {446, 521}

\bibitem[\protect\citeauthoryear{{Sijacki}, {Springel}, {Di Matteo}  \&
  {Hernquist}}{{Sijacki} et~al.}{2007}]{sijacki2007}
{Sijacki} D.,  {Springel} V.,  {Di Matteo} T.,   {Hernquist} L.,  2007, \mn@doi
  [\mnras] {10.1111/j.1365-2966.2007.12153.x}, \href
  {http://adsabs.harvard.edu/abs/2007MNRAS.380..877S} {380, 877}

\bibitem[\protect\citeauthoryear{{Sijacki}, {Vogelsberger}, {Genel},
  {Springel}, {Torrey}, {Snyder}, {Nelson}  \& {Hernquist}}{{Sijacki}
  et~al.}{2015}]{sijacki2015}
{Sijacki} D.,  {Vogelsberger} M.,  {Genel} S.,  {Springel} V.,  {Torrey} P.,
  {Snyder} G.~F.,  {Nelson} D.,   {Hernquist} L.,  2015, \mn@doi [\mnras]
  {10.1093/mnras/stv1340}, \href
  {http://adsabs.harvard.edu/abs/2015MNRAS.452..575S} {452, 575}

\bibitem[\protect\citeauthoryear{{Silk} \& {Rees}}{{Silk} \&
  {Rees}}{1998}]{silk1998}
{Silk} J.,  {Rees} M.~J.,  1998, \aap, \href
  {http://adsabs.harvard.edu/abs/1998A%26A...331L...1S} {331, L1}

\bibitem[\protect\citeauthoryear{{Tombesi}, {Sambruna}, {Reeves}, {Braito},
  {Ballo}, {Gofford}, {Cappi}  \& {Mushotzky}}{{Tombesi}
  et~al.}{2010}]{tombesi2010}
{Tombesi} F.,  {Sambruna} R.~M.,  {Reeves} J.~N.,  {Braito} V.,  {Ballo} L.,
  {Gofford} J.,  {Cappi} M.,   {Mushotzky} R.~F.,  2010, \mn@doi [\apj]
  {10.1088/0004-637X/719/1/700}, \href
  {http://adsabs.harvard.edu/abs/2010ApJ...719..700T} {719, 700}

\bibitem[\protect\citeauthoryear{{Tombesi}, {Mel{\'e}ndez}, {Veilleux},
  {Reeves}, {Gonz{\'a}lez-Alfonso}  \& {Reynolds}}{{Tombesi}
  et~al.}{2015}]{tombesi2015}
{Tombesi} F.,  {Mel{\'e}ndez} M.,  {Veilleux} S.,  {Reeves} J.~N.,
  {Gonz{\'a}lez-Alfonso} E.,   {Reynolds} C.~S.,  2015, \mn@doi [\nat]
  {10.1038/nature14261}, \href
  {http://adsabs.harvard.edu/abs/2015Natur.519..436T} {519, 436}

\bibitem[\protect\citeauthoryear{{Trakhtenbrot} \& {Netzer}}{{Trakhtenbrot} \&
  {Netzer}}{2012}]{trakhtenbrot2012}
{Trakhtenbrot} B.,  {Netzer} H.,  2012, \mn@doi [\mnras]
  {10.1111/j.1365-2966.2012.22056.x}, \href
  {http://adsabs.harvard.edu/abs/2012MNRAS.427.3081T} {427, 3081}

\bibitem[\protect\citeauthoryear{{Trakhtenbrot}, {Netzer}, {Lira}  \&
  {Shemmer}}{{Trakhtenbrot} et~al.}{2011}]{trakhtenbrot2011}
{Trakhtenbrot} B.,  {Netzer} H.,  {Lira} P.,   {Shemmer} O.,  2011, \mn@doi
  [\apj] {10.1088/0004-637X/730/1/7}, \href
  {http://adsabs.harvard.edu/abs/2011ApJ...730....7T} {730, 7}

\bibitem[\protect\citeauthoryear{{Trakhtenbrot}, {Volonteri}  \&
  {Natarajan}}{{Trakhtenbrot} et~al.}{2017}]{trakhtenbrot2017}
{Trakhtenbrot} B.,  {Volonteri} M.,   {Natarajan} P.,  2017, \mn@doi [\apjl]
  {10.3847/2041-8213/836/1/L1}, \href
  {http://adsabs.harvard.edu/abs/2017ApJ...836L...1T} {836, L1}

\bibitem[\protect\citeauthoryear{{Vogelsberger} et~al.,}{{Vogelsberger}
  et~al.}{2014}]{vogelsberger2014}
{Vogelsberger} M.,  et~al., 2014, \mn@doi [\mnras] {10.1093/mnras/stu1536},
  \href {http://adsabs.harvard.edu/abs/2014MNRAS.444.1518V} {444, 1518}

\bibitem[\protect\citeauthoryear{{Volonteri}, {Silk}  \& {Dubus}}{{Volonteri}
  et~al.}{2015}]{volonteri2015}
{Volonteri} M.,  {Silk} J.,   {Dubus} G.,  2015, \mn@doi [\apj]
  {10.1088/0004-637X/804/2/148}, \href
  {http://adsabs.harvard.edu/abs/2015ApJ...804..148V} {804, 148}

\bibitem[\protect\citeauthoryear{{Wagner}, {Umemura}  \& {Bicknell}}{{Wagner}
  et~al.}{2013}]{wagner2013}
{Wagner} A.~Y.,  {Umemura} M.,   {Bicknell} G.~V.,  2013, \mn@doi [\apjl]
  {10.1088/2041-8205/763/1/L18}, \href
  {http://adsabs.harvard.edu/abs/2013ApJ...763L..18W} {763, L18}

\bibitem[\protect\citeauthoryear{{Weinberger} et~al.,}{{Weinberger}
  et~al.}{2018}]{weinberg2018}
{Weinberger} R.,  et~al., 2018, \mn@doi [\mnras] {10.1093/mnras/sty1733}, \href
  {http://adsabs.harvard.edu/abs/2018MNRAS.479.4056W} {479, 4056}

\bibitem[\protect\citeauthoryear{{Williams}, {Maiolino}, {Krongold},
  {Carniani}, {Cresci}, {Mannucci}  \& {Marconi}}{{Williams}
  et~al.}{2016}]{williams2016}
{Williams} R.~J.,  {Maiolino} R.,  {Krongold} Y.,  {Carniani} S.,  {Cresci} G.,
   {Mannucci} F.,   {Marconi} A.,  2016, preprint, \href
  {http://adsabs.harvard.edu/abs/2016arXiv160508046W} {} (\mn@eprint {arXiv}
  {1605.08046})

\bibitem[\protect\citeauthoryear{{Willott} et~al.,}{{Willott}
  et~al.}{2010}]{willott2010}
{Willott} C.~J.,  et~al., 2010, \mn@doi [\aj] {10.1088/0004-6256/140/2/546},
  \href {http://cdsads.u-strasbg.fr/abs/2010AJ....140..546W} {140, 546}

\bibitem[\protect\citeauthoryear{{Zubovas} \& {King}}{{Zubovas} \&
  {King}}{2012}]{zubovas2012}
{Zubovas} K.,  {King} A.,  2012, \mn@doi [\apjl] {10.1088/2041-8205/745/2/L34},
  \href {http://adsabs.harvard.edu/abs/2012ApJ...745L..34Z} {745, L34}

\bibitem[\protect\citeauthoryear{{Zubovas}, {Nayakshin}, {Sazonov}  \&
  {Sunyaev}}{{Zubovas} et~al.}{2013}]{zubovas2013}
{Zubovas} K.,  {Nayakshin} S.,  {Sazonov} S.,   {Sunyaev} R.,  2013, \mn@doi
  [\mnras] {10.1093/mnras/stt214}, \href
  {http://adsabs.harvard.edu/abs/2013MNRAS.431..793Z} {431, 793}

\bibitem[\protect\citeauthoryear{{Zubovas}, {Bourne}  \& {Nayakshin}}{{Zubovas}
  et~al.}{2016}]{zubovas2016}
{Zubovas} K.,  {Bourne} M.~A.,   {Nayakshin} S.,  2016, \mn@doi [\mnras]
  {10.1093/mnras/stv2971}, \href
  {http://adsabs.harvard.edu/abs/2016MNRAS.457..496Z} {457, 496}

\makeatother
\end{thebibliography}

\appendix
\section{Wind propagation}
\label{app:wind}
In this appendix we derive analytic results and scaling laws for propagation of a AGN wind in a galactic medium. 
The interaction of the supersonic wind with the ambient medium inflates a shocked bubble which contains 
shocked wind material that crosses the reverse shock and  flows 
sideways, as well as shocked ambient gas that enters the bubble through the forward shock.  The
shocked wind and shocked ambient gas are separated by a contact discontinuity.   This structure, referred to
as cocoon in the preceding sections, is shown schematically in Fig. \ref{app:fig:scheme} and 
 is clearly visible in the snapshots of the density evolution (see, e.g., Fig \ref{fig:density_iso}).

An approximate, analytic calculation  of the wind dynamics 
employs momentum balance at the head (i.e., at the contact discontinuity).
Denoting the ambient gas and unshocked wind parameters by  subscripts $a$ and $w$, respectively, one finds:
\begin{equation}
\rho_w (v_w-v_h)^2+p_w=\rho_a (v_h-v_a)^2+p_a,
\end{equation}
where $v_a$ is the velocity of the ambient medium, $v_h$ the velocity of the head, and, henceforth, 
wind quantities ($\rho_w$ in particular) are measured just behind the reverse shock.   This 
result neglects gravitational forces.
In  the cases considered here, the wind moves against infalling matter, whereby $v_a$ is negative. 
Denoting $\alpha=\rho_a /\rho_w $, $a_a=\sqrt{p_a/\rho_a}$ and $a_w=\sqrt{p_w/\rho_w}$, we obtain
\begin{equation}
v_h=\frac{v_w-\alpha v_a}{1-\alpha}\left[1-\sqrt{1+(1-\alpha)\frac{\alpha v_a^2+\alpha a_a^2-v_w^2 - a_w^2}{(v_w-\alpha v_a)^2}}\right].
\end{equation}
If the wind is highly supersonic $v_w>> \alpha a_a, a_w$, then the solution for the head velocity simplifies to
\begin{equation}
v_h=\frac{v_w+\sqrt{\alpha}v_a}{1+\sqrt{\alpha}},
\label{app:eq:v_h_2}
\end{equation}
and wind propagation is possible provided $v_w > -\sqrt{\alpha} v_a$.   For the isothermal bulge considered in Sec. \ref{sec:results} 
we have to a good approximation $v_a\simeq -\sigma$ for the cold, free-falling gas (Eq. (\ref{app:eq:vel_ff})).  The later condition then
implies that the wind breaks out provided its velocity exceeds the escape velocity of the bulge (which is anyhow assumed by the neglect of 
the gravitational force).   Hence, $v_h = v_w/(1+\sqrt{\alpha})$ to a good approximation \citep[see also][]{begelman1984}. 
Now, in case of conical expansion $\rho_w\propto r^{-2}$, and if the ambient density scales as $\rho_a\propto r^{-p}$
then $\alpha\propto r^{2-p}$, which readily implies a constant head velocity if $p=2$, as in, e.g., the case with isothermal gas density
explored in Sec. \ref{sec:results}.   If, on the other hand, $p<2$ then $\alpha$ increases with radius and the cocoon decelerates.  
In fact, substantial deceleration is expected when the wind density becomes comparable to the ambient density, $\rho_w=\rho_a$.
For a wind with a total power $L_w=10^{46}L_{w46}$ erg s$^{-1}$ and opening angle $\theta_w$ this occurs at a radius
\begin{equation}
r_{dec} \simeq\left[\frac{10^4\, L_{w46}}{(1-\cos\theta_w)v_{w-1}^3 n_{a0}}\right]^{1/(2-p)}\quad {\rm pc}
\end{equation}
where $n_{a0}=\rho_{a0}/m_p$ is the number density of the ambient gas at a radius of 1 pc, in c.g.s. units.
For instance, for a BAL wind with $L_{w46}=1$, $v_w=10^4$ km s$^{-1}$ and $\theta_w=45^\circ$, 
expanding in a uniform density medium with $n_{a0}=1$, this gives $r_{dec}\sim 1$ kpc.
 Wind collimation may  alter this result.

\section{Temporal accretion profile for initially uniform gas distribution }
\label{app:sec:accretion}
For the spherical protogalaxy model invoked in section \ref{sec:results} the momentum equation reads:
\begin{equation}
\frac{d v_r}{dt}+\frac{1}{\rho_g}\frac{dp}{dr}=-\frac{GM_{BH}}{r^2}-\frac{Gm(r)}{r^2}=-\frac{\sigma^2}{r}\left(\frac{r_a}{r}+2\right),
\end{equation}
where $v_r$ is the radial velocity, $\rho_g$, $p$ are the gas density and pressure, respectively, and $r_a$ is the sphere of influence
defined in Eq. (\ref{eq:ra}).   Suppose that the gas density  is  uniform initially,  $\rho_g=$ const, and denote  $\tilde{r}=r/r_a$.  Then, in a 
hydrostatic equilibrium ($v_r=0$) the pressure profile is given by 
\begin{equation}
p(\tilde{r})=\rho_g\sigma^2\left[\frac{1}{\tilde{r}}-\frac{1}{\tilde{R}}+2\ln(\tilde{R}/\tilde{r})   \right] \equiv \rho_g\sigma^2 \Phi(\tilde{r}),
\label{app:eq:momentum}
\end{equation}
and satisfies $p(R)=0$, and the temperature profile by
\begin{equation}
T(\tilde{r})=\frac{m_p\sigma^2}{k} \Phi(\tilde{r}) \simeq 10^7\sigma^2_{300} \Phi(\tilde{r})\quad K.
\end{equation}

Now, if the gas is initially maintained at a hydrostatic equilibrium, it will quickly cool via free-free emission and inverse Compton (IC) scattering of the 
quasar radiation.   The free-free cooling time is given by 
\begin{equation}
t_{ff}\simeq10^7(\rho_g/m_p)^{-1}(T/10^7\, K)^{1/2}\quad yr.
\label{app:eq:ff_cooling}
\end{equation}
For a non-relativistic thermal electron distribution the inverse Compton cooling time can be expressed as
\begin{equation}
t_{c}=  \frac{3m_ec}{8\sigma_T u_{rad}}  \simeq 3\times10^3 \left(\frac{M_8 \tilde{r}^2}{ l \sigma_{300}^{4}}\right)   \quad yr
\label{app:eq:comp_cooling}
\end{equation}
in terms of the radiation energy density, $u_{rad} = L/4\pi c r^2$, where $L = l L_{Edd} = 10^{46} l M_8$ erg s$^{-1}$ is the 
quasar luminosity and $M_{BH}=10^8 M_8 M_\odot$ the SMBH mass.    
Consequently, free-free cooling dominates everywhere at densities $\rho_g/m_p >10^3$.  At much lower densities
it is naively expected that after a relatively short time the gas in the 
inner regions will be maintained at the Compton equilibrium temperature, while in the outer regions it will continue to cool via 
 free-free emission, although this might ultimately depend on accretion rate.   For instance, for our fiducial parameters in case B
we find $\dot{M}_{BH}/\dot{M}_{Edd}\simeq 6\times10^{-4}$, for which accretion is in the ADAF regime ($l << 10^{-3}$), so that
in this instance cooling is dominated by free-free emission everywhere.    In case of a relativistic electron distribution the cooling time is 
obtained from Eq. (\ref{app:eq:comp_cooling}) upon multiplication by the factor $m_ec^2/2kT$.  This is mainly relevant to the 
shocked wind gas at wind velocities $v_w > (m_e/m_p)^{1/2}c \simeq 0.02 c$.

\begin{figure}
\includegraphics[width=1\columnwidth]{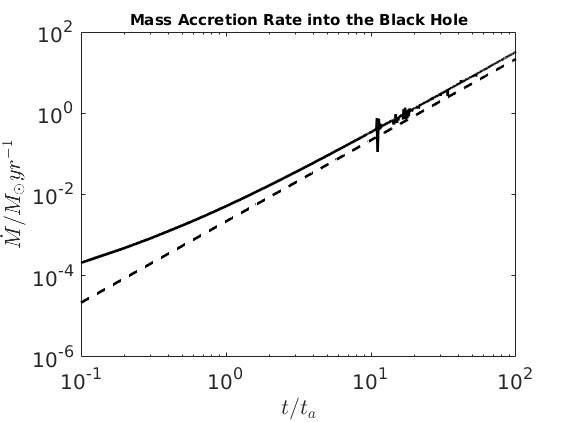}
\caption{Temporal evolution of  mass accretion rate in the absence of AGN feedback, for initial ambient 
density $\rho_0/m_p=1$.  The solid line delineates the result of
the 2D simulations. The dashed line is a plot of the analytic solution, Eq. (\ref{app:eq:Mdot_ev}).}
\label{app:fig:Mdot}
\end{figure}

Once the gas cools sufficiently, it starts accelerating and accretion into the SMBH gradually increase.
To determine the temporal accretion profile suppose for simplicity that the resting gas is initially cold ($kT < m_p\sigma^2$),
with a uniform density $\rho_g(t=0)=\rho_0$.
Neglecting the pressure in Eq. (\ref{app:eq:momentum}) readily yields
\begin{equation}
v_r(\tilde{r})=- \sigma \left[\frac{2}{\tilde{r}}-\frac{2}{\tilde{r}_0}+4\ln(\tilde{r}_0/\tilde{r})   \right]^{1/2},\quad \tilde{r} < \tilde{r}_0,
\label{app:eq:vel_ff}
\end{equation}
for a fluid element initially at rest at some radius $r_0$.  Outside the sphere of influence, $1< \tilde{r} < \tilde{r}_0$, the
free fall velocity is to a good approximation constant, $v_r\simeq -\sigma$.  The time it takes a fluid element, initially
at rest at a radius $r_0$, to reach this velocity is  $t\sim r_0/\sigma$.   Hence, the accretion front propagates roughly as $r(t)\sim \sigma t$.
Assuming a constant velocity $v_r=-\sigma$ within the accretion front ($r<\sigma t$) and $v_r=0$ outside ($r>\sigma t$), the 
solution to the continuity equation, $\partial_t \rho_g + r^{-2}\partial_r(r^2 v_r \rho_g)=0$, readily yields 
\begin{equation}
\rho_g(r,t) = \rho_0 (1+\sigma t/r)^2.
\end{equation}
At $r<< \sigma t$ we have to a good approximation $\rho_g =\rho_0 (\sigma t/r)^2$.    The associated mass accretion rate is 
\begin{equation}
\dot{M}(t) =4\pi r^2 \rho_{g}\sigma   \simeq 4\pi \rho_{0}\sigma r_a^2 ( t/t_a)^2.
\label{app:eq:Mdot_ev}
\end{equation}
A comparison between the analytical and numerical solution is exhibited in Fig. \ref{app:fig:Mdot}. 

\end{document}